\documentstyle[preprint,prb,aps]{revtex}

\begin{document}
\preprint{}
\title{Linear response theory of Coulomb drag in \\ coupled
electron systems}
\author{Karsten Flensberg\cite{byline1}, Ben Yu-Kuang Hu, Antti-Pekka Jauho}
\address{Mikroelektronik Centret, Danmarks Tekniske Universitet,
DK-2800 Lyngby, Denmark}
\author{Jari M. Kinaret\cite{byline2}}
\address{Nordita, Blegdamsvej 17, DK-2100 K\o benhavn \O, Denmark}
\date{21 April, 1995}
\maketitle
\begin{abstract}

We report a fully microscopic
theory for transconductivity, or, equivalently, momentum transfer rate,
of Coulomb coupled electron systems.  We use
the Kubo linear response formalism, and our main formal result expresses the
transconductivity in terms of two fluctuation diagrams, which are topologically
related, but not equivalent to, the Aslamazov-Larkin
diagrams known for superconductivity.  Previously reported results are
shown to be special cases of our general expression; specifically,
for constant impurity scattering rates, we recover the Boltzmann
equation results in the semiclassical clean limit, and the memory
function results in dirty systems.  Furthermore, we show that for energy
dependent relaxation times, the final result is not expressible
in terms of standard density-response functions.  Other new results include:
(i) at $T=0$, the frequency dependence of the transfer rate
is found to be proportional to $\Omega$ and $\Omega^2$ for frequencies
below and above the impurity scattering rate, respectively and (ii) the weak
localization correction to the transconductivity is given by
$\delta\sigma^{WL}_{21} \propto \delta\sigma^{WL}_{11}
+\delta\sigma^{WL}_{22}$.

\end{abstract}

\narrowtext
\section{Introduction}
\label{sec:intro}
Consider two systems containing mobile charge carriers
so close to each other that the charges in the
two respective subsystems feel the Coulomb forces originating
from the other subsystem, and yet far enough away from each other
that direct charge transfer between the two subsystems is not
possible.  Experimental realizations of such systems are,
for example, Coulomb coupled double quantum well systems,\cite{gram_all,siva92}
arrangements where a 3D system is close
to a 2D system,\cite{solo89} or two nearby quantum wires.
A scattering event between  a carrier in one system
and a carrier in the other
system leads to momentum
transfer between the two subsystems.
Thus, if a current is driven through one of the systems (henceforth
the driven system is denoted as layer 1), then an induced
current is dragged in the other subsystem (layer 2).
Alternatively,
if no current is allowed to flow in layer 2, a
voltage is induced.
Due to momentum conservation  the two particle number currents flow in
the same direction. Since the mechanism for the Coulomb drag is
carrier-carrier scattering the drag current is proportional to
the square of the effective interaction between the subsystems.
The available phase space for electron-electron scattering
tends to zero
at low temperatures, and consequently
one expects Coulomb drag to decrease with
decreasing temperature. At low temperatures, the two Pauli
factors entering the carrier--carrier
scattering rate lead to a $T^2$-dependence, and
this behavior is approximately seen in experiments.\cite{gram_all}
Note, however, that there are small, but important deviations
from the simple $T^2$-law; these deviations have been the
topic of much recent interest.\cite{gram_all,tso,flen94}

The possibility for Coulomb drag was realized already long
ago,\cite{pogr77,pric83} and the recent experimental advances
\cite{gram_all,siva92,solo89} have brought about a flurry
of theoretical works.  A number of different theoretical
approaches has been proposed.
These include (i) calculations based on the Boltzmann
equation,\cite{gram_all,jauh93} (ii) the memory function
approach of Ref.\onlinecite{zhen93}, and
(iii) the momentum balance equation method.\cite{tso}

In this paper, we calculate the Coulomb drag between
two systems using a fully microscopic theory based on a linear
response formula.  The central object to be evaluated
is the retarded current-current correlation function;
since the two involved currents refer to the the
two different subsystems, we call the result of
this calculation {\it transconductivity}.
The motivation underlying our work is that all previously
proposed approaches lack the rigor that can be
achieved with a formal linear-response calculation.
The present method allows us to identify the Feynman
diagrams that contribute to the transconductivity.  Instead
of the normal conductivity bubble, we find that one
must evaluate a fluctuation diagram, which is similar,
but not identical, to the Aslamazov-Larkin\cite{asla68}
diagrams known from superconductivity, or the diagrams
encountered in connection with the microscopic theory
of van der Waals interactions.\cite{rapc91,gold89}  Thus, all
the methods developed within the diagrammatic perturbation
theory are readily applicable, and one can systematically study
the effects of higher order scattering processes, such
as vertex corrections or weak localization, or electron-phonon
interactions, or the effect of magnetic fields.

Apart from the general formulation for calculating
transconductivities, we obtain the following
explicit results.
In the limit of weak impurity scattering we show that the
linear-response result reduces to the expression obtained with the
Boltzmann equation. We also study the corrections to the
Boltzmann equation formula in the case of stronger impurity scattering,
{\it i.e.}\ accounting for vertex corrections.
Further, we show that the Boltzmann equation result, which involves
the susceptibility functions of the individual subsystems, must
be generalized, if one considers energy-dependent scattering rates.
We also consider weak localization corrections to the
transconductivity.  All these effects are calculated
at finite temperature, but for zero external frequency.
At $T=0$ finite frequency calculations become feasible, and
we present a general proof that the dc-drag current vanishes
in the dc limit at zero temperature  for an {\it open} system.
Recently, Rojo and Mahan\cite{rojo92}
used a ground state energy argument to calculate
the drag current, and found a nonzero result at zero
temperature.  This seems to contradict our results, however
there is an important distinction: the calculation of
Ref.\ \onlinecite{rojo92} applies
for {\it closed} systems, {\it e.g.} coupled mesoscopic rings.

This paper is organized as follows.  Section \ref{sec:basics} outlines
the derivation
of the general expression for transconductivity.
Section \ref{sec:imp} is devoted to impurity scattering: first, we show how
the well-known Boltzmann result follows from the general formulation,
and next, we establish a connection to the memory function formulation
of Ref.\ \onlinecite{zhen93}. This section is concludes with a
discussion of energy-dependent scattering rates and weak
localization effects.  Section \ref{sec:zero}
presents result for $T=0$ at finite frequency.  Finally, a
number of technical details can be found in the appendices.

\section{General formula for the transconductivity}
\label{sec:basics}
The previous works have related Coulomb drag to the
{\it transresistivity}, $\rho_{21}$; our calculation, which is based on a
Kubo formula leads to the {\it transconductivity}, $\sigma_{21}$.
These are defined as
\begin{mathletters}
\begin{eqnarray}
\rho_{21} &=& \frac{E_2}{J_1},\ \ \mbox{with}\ J_2 = 0;\\
\sigma_{21} &=& \frac{J_2}{E_1},\ \ \mbox{with}\ E_2 = 0,
\end{eqnarray}
\end{mathletters}
where $E_i$ and $J_i$ are, respectively,
the electric field and the current density in layer $i$.
These two quantities are related via
\begin{equation}
\rho_{21} = \frac{-\sigma_{21}}{\sigma_{11}\sigma_{22}-\sigma_{12}\sigma_{21}}
\approx  \frac{-\sigma_{21}}{\sigma_{11}\sigma_{22}}\;
\label{transres}
\end{equation}
In (\ref{transres}) the diagonal $\sigma$'s are the individual
subsystem conductivities and we note that the transconductivity is always much
smaller than intralayer conductivity, because it is caused
by a screened interaction between spatially separated systems
({\it e.g.} the data of Ref.\ \onlinecite{gram_all}  gives $\sigma_{21}/
\sigma_{11}\simeq 10^{-6}$).
The transresistivity $\rho_{21}$ is a more physically relevant quantity than
$\sigma_{21}$ in a drag-rate measurement because $\rho_{21}$
is directly related to
the rate of momentum transfer from particles in layer 1 to layer 2,
$\tau_{21}^{-1}$, without reference to the
scattering rates of the individual layers; i.e.,
\begin{equation}
\rho_{21} = \frac{m_1}{n_1 e^2 \tau_{21}};\qquad\qquad
\frac{1}{\tau_{21}} = \frac
{(\overline{{\partial p_2}/{\partial t }})_{\mathrm{drag}}}
{\overline{p}_1},
\end{equation}
where $m$ is the effective mass, $n$ is the carrier density,
$p$ is the momentum per particle, $(\partial p/\partial t)_{\mathrm{drag}}$
is the momentum transfer due to interlayer interactions,
and the overline denotes an ensemble average.

The Kubo formula\cite{mahan}
expresses the conductivity tensor in terms of the
retarded current-current
correlation function,
\begin{equation}
\sigma_{ij}^{\alpha\beta}({\mbox{\boldmath $x$}}-{\mbox{\boldmath
$x'$}};\Omega) =
\frac{ie^2}{\Omega}\Pi^{\alpha\beta,r}_{ij}({\mbox{\boldmath $x$}}-
{\mbox{\boldmath $x'$}};\Omega)
+ \frac{ie^2}{m\Omega}\delta({\mbox{\boldmath $x$}} - {\mbox{\boldmath
$x'$}})\delta_{ij}
\delta_{\alpha\beta}\rho_i({\mbox{\boldmath $x$}}),
\end{equation}
where (throughout we use $\hbar=1$)
\begin{equation}\label{PiHeis}
\Pi^{\alpha\beta,r}_{ij}({\mbox{\boldmath $x$}}-{\mbox{\boldmath $x'$}};t-t') =
-i \Theta(t-t') \langle
[ j_i^{\alpha}(\mbox{\boldmath $x$},t), j_j^{\beta}
(\mbox{\boldmath $x'$},t')]\rangle.
\end{equation}
Here $\{ij\}$ indicate the subsystem, $\{\alpha\beta\}$ in the
superscripts label the Cartesian coordinates,
$\rho_i(\mbox{\boldmath $x$})$ is the particle density in subsystem $i$,
and $\mbox{\boldmath $j$}(\mbox{\boldmath $x$},t)$ is the
particle current operator.  We have  assumed that the subsystems
are translationally invariant.  Our task consists of calculating the
transconductivity $\sigma_{21}^{\alpha\beta}$.

We employ the imaginary-time formalism to evaluate the retarded
current-current correlation function, starting with the
(imaginary-)time-ordered correlation function
\begin{equation}
\Pi^{\alpha\beta}_{21}(\mbox{\boldmath $x$}-\mbox{\boldmath $x'$};\tau-\tau') =
-\langle
 T_\tau\{ j_1^{\alpha}(\mbox{\boldmath $x$},\tau) j_2^{\beta}(\mbox{\boldmath
$x'$},\tau')\}\rangle.
\end{equation}
The retarded function then follows as
\begin{equation}\label{ret}
\Pi^{\alpha\beta,r}_{21}(\mbox{\boldmath $x$}-\mbox{\boldmath $x'$};\Omega) =
\lim_{i\Omega_n\rightarrow \Omega +i\delta}
\Pi_{21}^{\alpha\beta}(\mbox{\boldmath $x$}-\mbox{\boldmath $x'$};i\Omega_n),
\end{equation}
where
\begin{eqnarray}
\Pi_{21}^{\alpha\beta}(\mbox{\boldmath $x$}-\mbox{\boldmath $x'$};i\Omega_n) &
= &
\int_0^\beta d\tau \,
e^{i\Omega_n \tau}\Pi_{21}^{\alpha\beta}(\mbox{\boldmath $x$}-\mbox{\boldmath
$x'$};\tau)\\
\Pi_{21}^{\alpha\beta}(\mbox{\boldmath $x$}-\mbox{\boldmath $x'$};\tau) & = &
{1\over\beta}
\sum_n e^{-i\Omega_n\tau}
\Pi_{21}^{\alpha\beta}(\mbox{\boldmath $x$}-\mbox{\boldmath $x'$};i\Omega_n),
\end{eqnarray}
and where $\beta = 1/k_B T$.
The calculation proceeds by expanding the transconductivity in powers
of the interaction between the
subsystems.
The interaction Hamiltonian is given by
\begin{equation}
H_{12} =\int d\mbox{\boldmath $r_1$}\int d\mbox{\boldmath $r_2$}\
\rho_1(\mbox{\boldmath $r_1$})\,
U_{12}(\mbox{\boldmath $r_1$} - \mbox{\boldmath $r_2$}) \,
\rho_2(\mbox{\boldmath $r_2$}).
\end{equation}
Here $U_{12}$ is the bare Coulomb interaction between the
the systems.
We note that other interaction processes, which couple the
charge carriers in the two subsystems, can be treated similarly.
An important example is the virtual phonon mediated interaction,
which may play a role in the low temperature behavior of the
momentum transfer rate.\cite{gram_all,tso}

The $\tau$-dependence of the current operators in (\ref{PiHeis})
is determined
by the full Hamiltonian $H = H_1 + H_2 + H_{12}$, where $H_i$ are
the subsystem Hamiltonians.
In order to develop a perturbation expansion, we
must isolate the $H_{12}$-dependence.  Following the
standard many-body prescription,\cite{mahan} we
transform into the interaction representation and obtain
\begin{equation}\label{piint}
\Pi_{21}^{\alpha\beta}(\mbox{\boldmath $x$} - \mbox{\boldmath
$x'$},\tau-\tau')=
-\langle T_{\tau}\{ S(\beta) {\hat j}_1^{\alpha}(\mbox{\boldmath $x$},\tau)
\hat j_2^{\beta}(\mbox{\boldmath $x'$},\tau') \} \rangle ,
\end{equation}
where the carets indicate that the time-dependence is now governed
by the individual subsystem Hamiltonians, and the operator $S(\beta)$ is
\begin{equation}\label{Sm}
S(\beta)=T_{\tau}\Bigl\{\exp[-\int_0^{\beta}d\tau_1\hat H_{12}(\tau_1)]\Bigr\}.
\end{equation}
As usual, only connected diagrams need to be included.
It is now straightforward to expand $S(\beta)$ in powers of
$\hat H_{12}$,
\begin{eqnarray}\label{Sseries}
S(\beta) \approx 1 &-& T_{\tau}\Bigl\{\int_0^{\beta}d\tau_1\hat H_{12}(\tau_1)
\Bigr\}\nonumber\\
&+&\frac{1}{2}T_{\tau}\Bigl\{\int_0^{\beta}d\tau_1\int_0^{\beta}d\tau_2
\hat H_{12}(\tau_1) \hat H_{12}(\tau_2)\Bigr\} + \dots
\end{eqnarray}
The zeroth order term leads to a vanishing contribution to
the transconductivity because the two current operators
are decoupled and hence commute. In the following sections
we discuss the higher order terms.

\subsection{Linear expansion}

The linear order term in $H_{12}$
leads to the correlation function
\begin{eqnarray}
\Pi_{21}^{\alpha\beta}(\mbox{\boldmath $x$}-\mbox{\boldmath $x'$},\tau -
\tau')^{(1)} &=&
\int_0^{\beta}d\tau_1\int d\mbox{\boldmath $r_1$}  d\mbox{\boldmath $r_2$}
\langle T_{\tau}\{ \hat j^{\alpha}_1(\mbox{\boldmath $x$},\tau)
\hat\rho_1(\mbox{\boldmath $r_1$},\tau_1) \}\rangle\nonumber\\
&\times&U_{12}(\mbox{\boldmath $r_1$}-\mbox{\boldmath $r_2$})
\langle T_{\tau} \{ \hat\rho_2(\mbox{\boldmath $r_2$},\tau_1)
\hat j^{\beta}_2(\mbox{\boldmath $x'$},\tau') \} \rangle.
\end{eqnarray}
Use of the continuity equation, $i\Omega\rho+\nabla\cdot
\mbox{\boldmath $j$}=0$, allows
us to eliminate the number density operator, and to
express the density--current correlators in terms of the subsystem
conductivities.  After some simplification we find (for
a translationally invariant impurity averaged system
where impurity scattering in the two subsystems is
uncorrelated\cite{non-trans})
\begin{equation}\label{1storder}
\sigma_{21}^{\alpha\beta}(\mbox{\boldmath $q$},\Omega)^{(1)} =
{1\over ie^2\Omega} \sum_{\gamma,\delta}
\sigma_{22}^{\alpha\gamma}(\mbox{\boldmath $q$},\Omega)\, q^\gamma\,
U_{12}(\mbox{\boldmath $q$})\, q^\delta \,
\sigma^{\delta\beta}_{11}(\mbox{\boldmath $q$},\Omega).
\end{equation}
This expression is exact, and it can be used to calculate
the first order transconductivity for any system, once
the subsystem conductivities $\sigma_{ii}$ are known.\cite{6-lines}
{}From (\ref{1storder}) we also infer that the
first-order transconductivity vanishes in the dc-limit.

\subsection{Quadratic expansion}

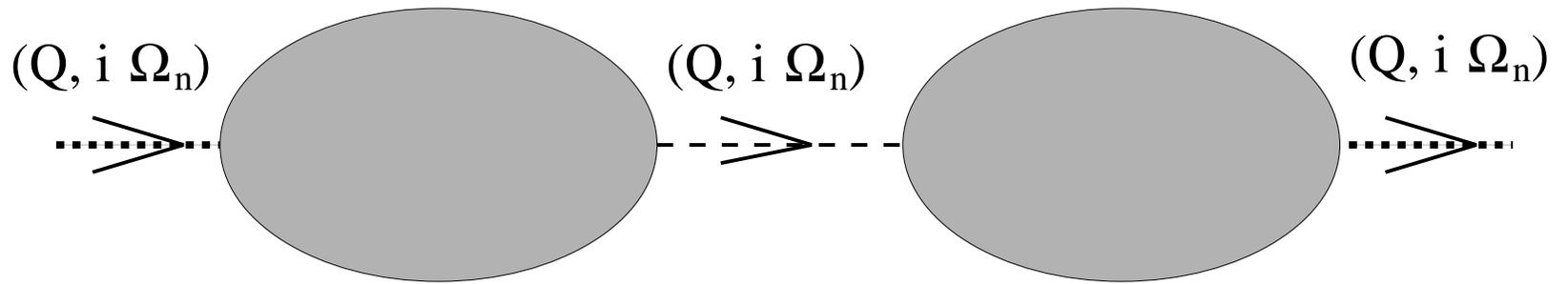
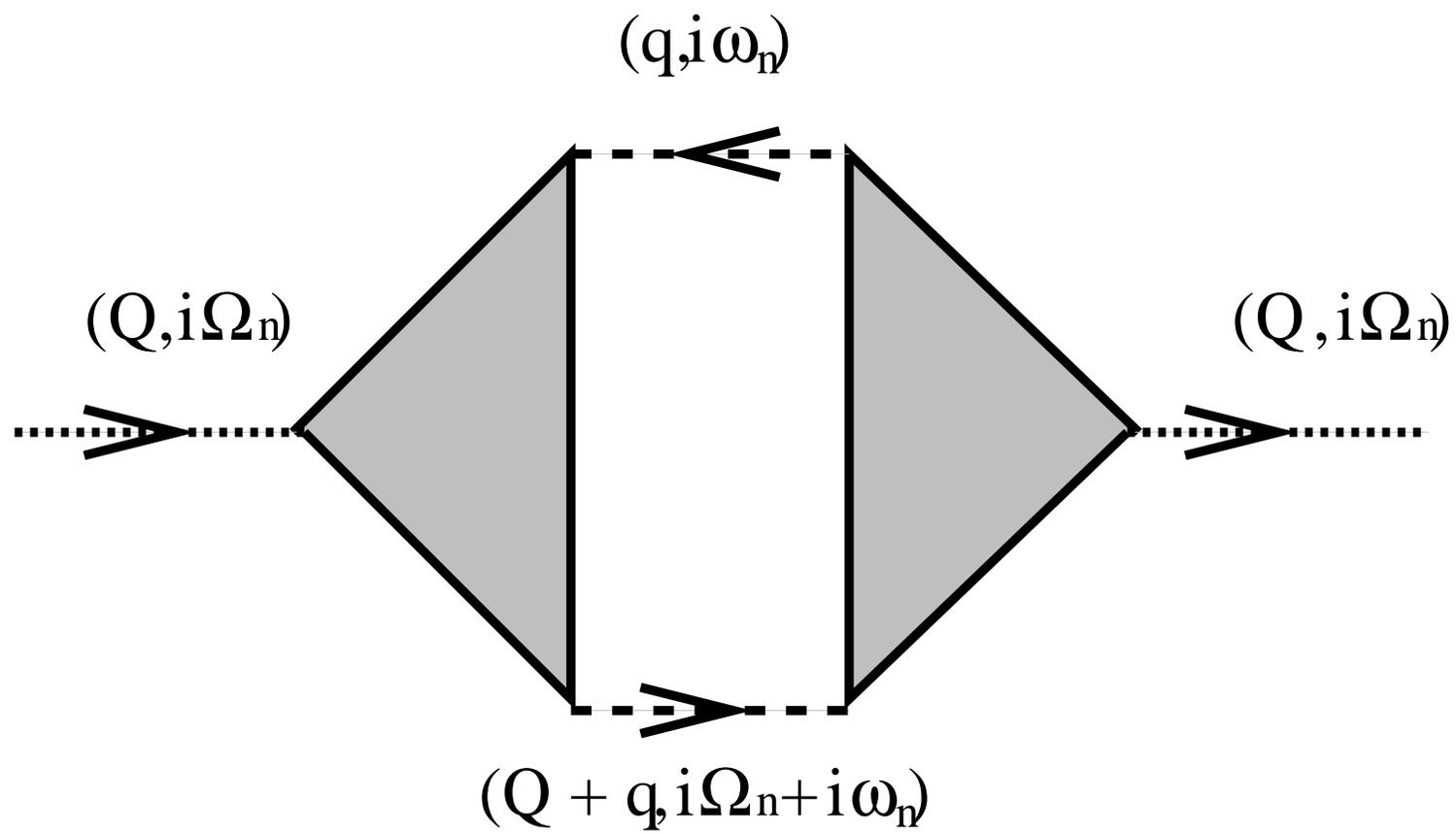
\begin{figure}
\caption{Diagrams corresponding to the current--current correlation function,
to (a) first and (b) second order in the interlayer Coulomb interaction.
The shaded triangles correspond to the $\Delta$s
given in Eqs.\ (\protect{\ref{Deltadef}}) and
(\protect{\ref{Deltafourierdef}}), the dashed lines to the interaction,
the dotted lines to the external current operators, and the arrowheads
to the direction of momentum and energy transfer.}
\label{fig:diagram}
\end{figure}

To evaluate $\Pi_{21}^{\alpha\beta}$ to second order in $H_{12}$,
we substitute the third term of (\ref{Sseries})
on the right-hand side of (\ref{piint}),
and find that the current-current
correlation function is given by
\begin{eqnarray}\label{2deltas}
\Pi_{21}^{\alpha\beta}(\mbox{\boldmath $x$} -
\mbox{\boldmath $x'$};\tau-\tau')^{(2)} &=&
-\frac{1}{2} \int_0^\beta d\tau_1
\int_0^\beta d\tau_2
\int d\mbox{\boldmath $r_1$} \int  d\mbox{\boldmath $r_2$}
\int d\mbox{\boldmath $r_1'$} \int d\mbox{\boldmath $r_2'$} \, \nonumber\\
&\times&U_{12}(\mbox{\boldmath $r_1$}-\mbox{\boldmath $r_2$})
U_{12}(\mbox{\boldmath $r_1'$}-\mbox{\boldmath $r_2'$})
\Delta_1^{\alpha}(\mbox{\boldmath $x$}\tau,\mbox{\boldmath
$r_1$}\tau_1,\mbox{\boldmath $r_1'$}\tau_2)
\Delta_2^{\beta}(\mbox{\boldmath $x'$}\tau',\mbox{\boldmath
$r_2$}\tau_1,\mbox{\boldmath $r_2'$}\tau_2),
\end{eqnarray}
where we have defined the function
\begin{equation}\label{Deltadef}
\Delta_i^{\alpha}(\mbox{\boldmath $x$}\tau,\mbox{\boldmath $x'$}\tau',
\mbox{\boldmath $x''$}\tau'') =
-\langle
T_\tau\{ \hat j_i^{\alpha}(\mbox{\boldmath $x$}\tau) \hat\rho_i(\mbox{\boldmath
$x'$}\tau')
\hat\rho_i(\mbox{\boldmath $x''$}\tau'')\}\rangle.
\end{equation}
Just as in the previous section,
we factorized the time-ordered expectation value
involving two current and four density
operators; this step is justified because the two subsystems
are decoupled after the formal expansion in $H_{12}$.
Due to the assumed translational invariance $\Delta$
depends on only two coordinate differences.  We
define
the Fourier transform
$\Delta({\mbox{\boldmath $q$}},{\mbox{\boldmath $q'$}};\omega,\omega')$
via ($\nu=$ the volume)
\begin{eqnarray}\label{Deltafourierdef}
\Delta(\mbox{\boldmath $x$}\tau,\mbox{\boldmath $x'$}\tau',
\mbox{\boldmath $x''$}\tau'') =
\frac{1}{\nu^2}\sum_{\mbox{\boldmath $q_1$},\mbox{\boldmath $q_2$}}
\frac{1}{\beta^2}\sum_{i\omega_m,i\omega_n}
&& e^{i\mbox{\boldmath $q_1$}\cdot(\mbox{\boldmath $x$}-\mbox{\boldmath $x''$})
+
i\mbox{\boldmath $q_2$}\cdot (\mbox{\boldmath $x'$}-\mbox{\boldmath $x''$})}
e^{-i\omega_m(\tau-\tau'') - i\omega_n(\tau'-\tau'')}\nonumber\\
&&\times\Delta(\mbox{\boldmath $q_1$}+\mbox{\boldmath $q_2$},\mbox{\boldmath
$q_2$};i\omega_m+i\omega_n,i\omega_n).
\end{eqnarray}
The final expression for $\Pi_{21}^{\alpha\beta}(\mbox{\boldmath $Q$},
i\Omega_n)^{(2)}$ then takes the form
\begin{eqnarray}\label{PiDelta}
\Pi_{21}^{\alpha\beta}(\mbox{\boldmath $Q$};i\Omega_n)^{(2)} &=& -\frac{1}{2}
\frac{1}{\nu}\sum_{\mbox{\boldmath $q$}} \frac{1}{\beta}\sum_{i\omega_n}
U_{12}(\mbox{\boldmath $q$})U_{12}^{*}(\mbox{\boldmath $Q$}+\mbox{\boldmath
$q$})\nonumber\\
&&\Delta_1^{\alpha}(\mbox{\boldmath $Q$}+\mbox{\boldmath $q$},\mbox{\boldmath
$q$};i\Omega_n+i\omega_n,i\omega_n)
\Delta_2^{\beta}(-\mbox{\boldmath $Q$}-\mbox{\boldmath $q$},-\mbox{\boldmath
$q$};-i\Omega_n-i\omega_n,-i\omega_n).
\end{eqnarray}
The diagram corresponding to the second order result, Eq.(\ref{PiDelta}),
is shown in Fig.\  \ref{fig:diagram}.
We also display the first order term discussed above.

Consider next the $i\omega_n$ summations. The function
$\Delta(z+i\Omega_n,z)$
has branch cuts in the complex $z$-plane at
Im($i\Omega_n+z$)=0, or Im($z$)=0, and is analytic elsewhere
(see Appendix \ref{app:branchcuts}).
We can therefore perform the $i\omega_n$-sum as a contour integral.
When we extract the retarded part according to Eq. (\ref{ret}), we
obtain the result
\begin{mathletters}
\begin{eqnarray}\label{S}
S(\Omega)&\equiv&
\frac{1}{\beta}\left. \sum_{i\omega_n}
\Delta_1(i\Omega_n+i\omega_n,i\omega_n)
\Delta_2(-i\Omega_n-i\omega_n,-i\omega_n)
\right|_{i\Omega_n\rightarrow\Omega+i\delta}\equiv S_1(\Omega)+S_2(\Omega),\\
S_1(\Omega)&=&
{\mbox{$\cal P$}}\int_{-\infty}^\infty \frac{d\omega}{2\pi i}
\left[n_{\mathrm B}(\omega+\Omega) -n_{\mathrm B}(\omega)\right]
\Delta_1(+,-)\Delta_2(-,+)\label{S1}\\
S_2(\Omega)&=&
{\mbox{$\cal P$}}\int_{-\infty}^\infty \frac{d\omega}{2\pi i}\left[
n_{\mathrm B}(\omega)
\Delta_1(+,+)\Delta_2(-,-)
-n_{\mathrm B}(\omega+\Omega)\Delta_1(-,-)
\Delta_2(+,+)\right],\label{S2}
\end{eqnarray}
\end{mathletters}
where $n_{\mathrm{B}}(\omega)$
is the Bose function. We have suppressed the Cartesian indices and
the momentum labels, since they
can be gleaned from Eq.(\ref{PiDelta}), and have used the notation
\begin{mathletters}
\begin{eqnarray}\label{Deltapmdef}
\Delta_1(\pm,\pm) &=& \Delta(i\Omega_n+i\omega_n\rightarrow
\Omega+\omega{\pm}i\delta;\,i\omega_n\rightarrow \omega{\pm}i\delta),
\\
\Delta_2(\pm,\pm) &=& \Delta(-i\Omega_n-i\omega_n\rightarrow
-\Omega-\omega{\pm}i\delta,\,-i\omega_n\rightarrow -\omega{\pm}i\delta)
\;.
\end{eqnarray}
\end{mathletters}
The functions $\Delta_i(+,+)$ and $\Delta_i(-,-)$ vanish identically
in the dc limit, which is proven in Appendix \ref{app:zero}.

Consider next the dc-response, $\Omega\to 0$, of
a uniform system (${\mbox{\boldmath $Q$}}=0$).
The dc-limit of $S_1(\Omega)$ is simple to evaluate,
because the difference of the two Bose functions combined
the prefactor $\Omega^{-1}$
just gives $\partial_{\omega}n_{\mathrm{B}}(\omega)$.
Thus the dc-transconductivity reduces to
(reintroducing $\hbar$)
\begin{eqnarray}
\sigma_{21}^{\alpha\beta(2)}&=&\frac{e^2}{\hbar}
\Bigl(-{1\over 2}\Bigr)
\frac{1}{\nu}\sum_{\mbox{\boldmath $q$}}
\int_{-\infty}^\infty \frac{d\omega}{2\pi} \,
\left|U_{12}(q)\right|^2
[\partial_{\omega}n_{\mathrm B}(\omega)]\nonumber\\
&&\times
\Delta^{\alpha}_{1}(\mbox{\boldmath $q$},\mbox{\boldmath
$q$};\omega+i\delta,\omega-i\delta)
\Delta^{\beta}_{2}(-\mbox{\boldmath $q$},-\mbox{\boldmath
$q$};-\omega-i\delta,-\omega+i\delta)
\;.
\label{sigma21}
\end{eqnarray}
The actual evaluation of this expression at various levels of
approximation forms the main task
of this paper.  In the case of electron-hole systems the overall
sign of (\ref{sigma21}) must be changed.

\subsection{Higher order terms}

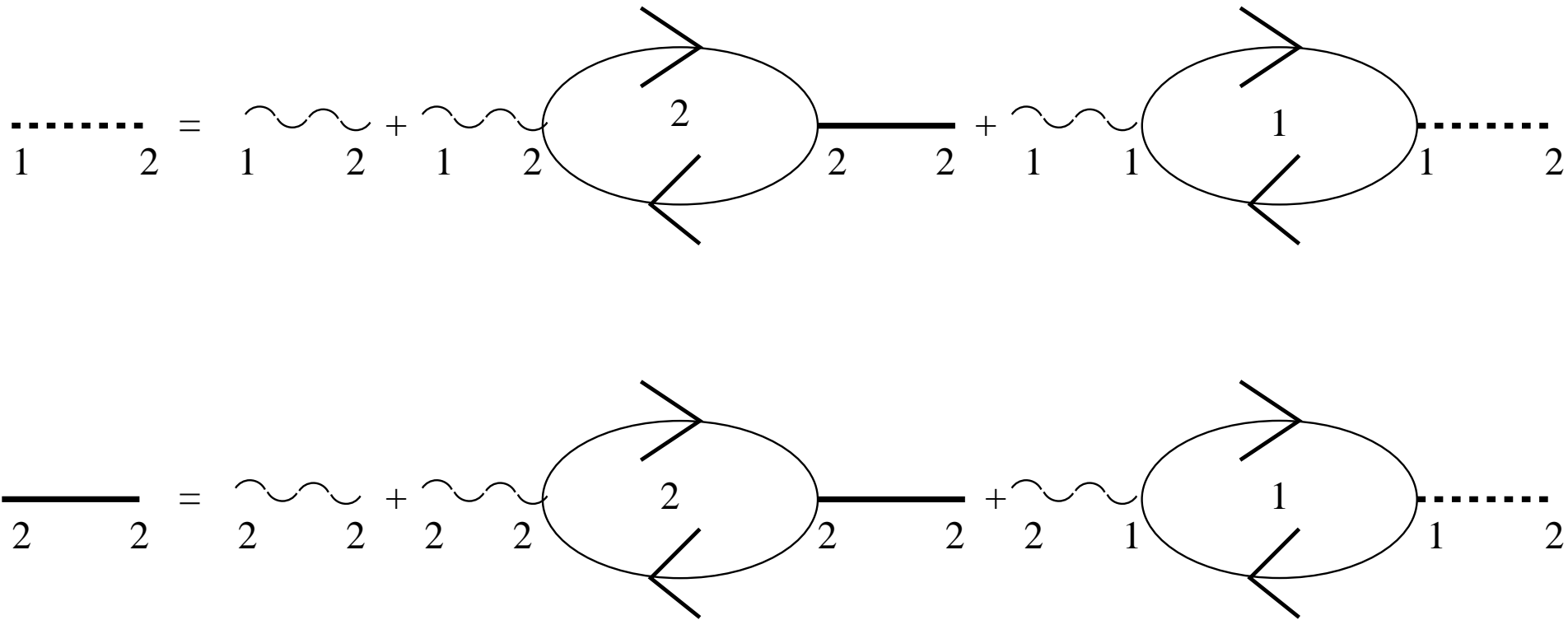
\begin{figure}
\caption{Diagrams which lead to the screened interlayer interaction
within the random phase approximation.  The ``bubbles" are the bare
polarizabilities of the subsystems, the thin wavy lines are the bare
interactions, the thick lines are the screened interactions, and
numbers indicate the subsystem.}
\label{fig:rpa}
\end{figure}

The $S$-matrix expansion (\ref{Sseries}) can be used to generate higher
order terms.  To proceed systematically
one must apply the techniques of the many-body formalism.
As usual, the most important processes should
be identified, and the corresponding diagrams be
summed to infinite order.  This procedure may then
lead to an integral equation for the effective
interaction, {\it e.g.} in the ladder approximation
one obtains the Bethe-Salpeter equation.
We do not pursue this line of argument further in this paper, but
note that a particularly useful resummation can
be obtained, if one includes the ``bubble"-diagrams
(see Fig.\ \ref{fig:rpa}),
which leads to an effective screened interaction,
$U_{12}(q)\to U_{12}(q,\omega) = U_{12}(q)/\epsilon_{12}(q,\omega)$,
where the dielectric function is given by
\begin{equation}\label{eps12}
\epsilon_{12}(q,\omega)=
[1-U_1(q)\chi_1(q,\omega)][1-U_2(q)\chi_2(q,\omega)]
-U_{12}(q)^2\chi_1(q,\omega)\chi_2(q,\omega) ,
\end{equation}
where the $\chi$'s are the usual polarization functions, and the $U_i$'s
are the intrasystem Coulomb interactions. We observe
that an energy-dependent $U(q,\omega)$ can be used
in the above expressions, (\ref{1storder}) and
(\ref{sigma21}), for transconductivity with no
additional difficulty.
Most previous works \cite{siva92,tso,flen94,jauh93,laik90}
on drag problems have used (\ref{eps12}) (or simplified versions of it).

\section{Impurity scattering}
\label{sec:imp}
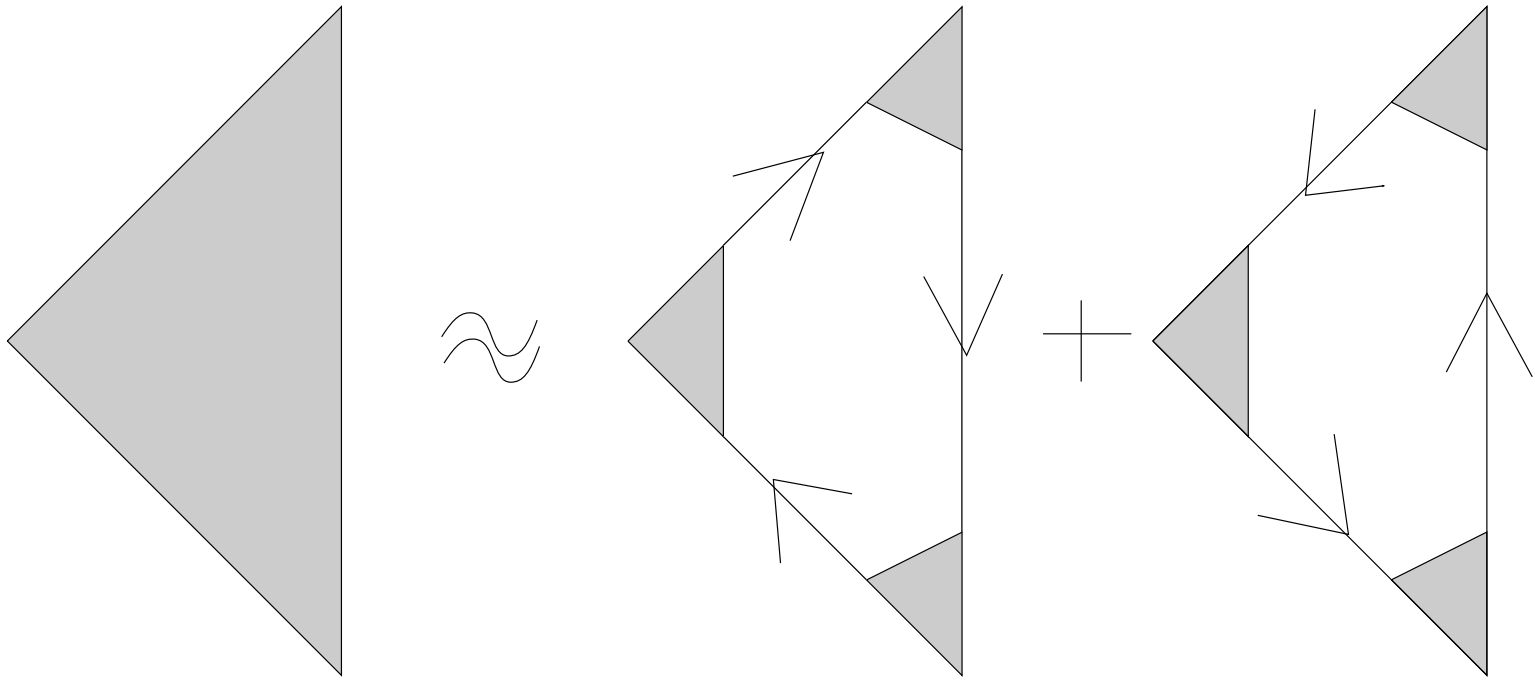
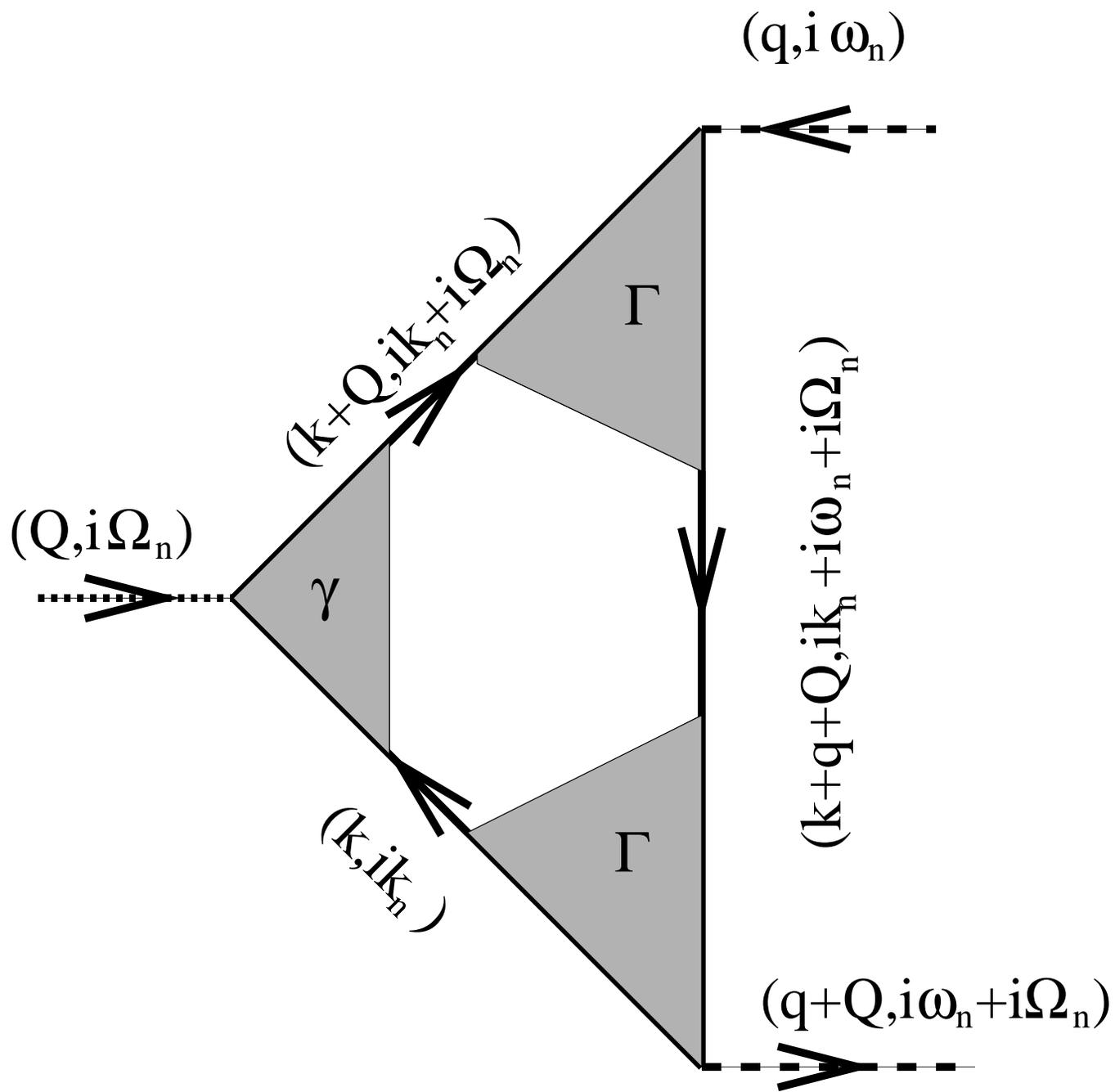
\begin{figure}
\caption{The function $\Delta$ for the case in which vertex corrections
are included at each of the individual charge and current vertices.
Figure (a) shows the decomposition of $\Delta$ into diagrams with
clockwise- and anticlockwise-moving Green functions, with the
grey shaded areas indicating vertex corrections.
Figure (b) shows one of these diagrams in greater detail.
Here $k^{\alpha}\gamma$ is the current vertex, the $\Gamma$ is
the charge vertex, the dashed
line is incoming momentum and frequency, the dotted lines are the
interaction $U_{12}(q,\omega)$, and the solid lines with arrows are the
Green function.  Normal momentum and energy conservation rules
apply at the vertices.}
\label{fig:vertex}
\end{figure}

In the previous section we showed that the transconductivity
can be expressed in terms of the general
three-body correlation function $\Delta$.
We will next consider a specific example in order to calculate
this three-body function, namely non-interacting electrons scattering
against random impurities.  The Hamiltonian representing impurity scattering
is quadratic, and hence Wick's theorem is applicable,
which means that the expectation value can be factorized into
pairwise contractions, {\it i.e.} expressed in terms of Green
functions.  Impurity averaging, which is now implicit in the
expectation value,
reintroduces correlations between the particles, which
implies that one must introduce vertex functions.
However, we do not allow impurity correlations between
the two subsystems, {\it i.e.} we assume that
$\langle \Delta_1 \Delta_2\rangle_{\mathrm{imp}} =
\langle \Delta_1\rangle_{\mathrm{imp}}\langle\Delta_2\rangle_{\mathrm{imp}}$.
\cite{imp-imp}
The particular choice for the impurity self-energy used
in the calculation of the impurity averaged Green function
fixes the choice of the vertex function; in what
follows we use the self-consistent Born approximation for
the self-energy, and the corresponding vertex function
consists either of the ladder diagrams (Section \ref{sec:ladder}), or of the
maximally crossed diagrams (Section \ref{sec:WL}).
The form for impurity-$\Delta$ giving the dominant contribution
is shown in Fig.\ \ref{fig:vertex}.\cite{crossed}
We consider only uniform systems, and set the
external wave vector to zero, ${\mbox{\boldmath $Q$}}=0$.
We also denote fermionic complex frequencies
by $ik_m$ in contrast to bosonic frequencies $i\omega_n$, and the
external frequency $\Omega$. Thus we have
\begin{eqnarray}
\Delta^{\alpha}({\mbox{\boldmath $q$}},{\mbox{\boldmath $q$}};i\Omega_n+ &&
i\omega_n,i\omega_n)
=-\frac{2}{m\nu} \sum_{\mbox{\boldmath $k$}} \frac{1}{\beta}\sum_{ik_m}
k^{\alpha}\Bigl[ {\mbox{$\cal K$}}({\mbox{\boldmath $k$}},{\mbox{\boldmath
$q$}}
,ik_m,i\Omega_n,i\omega_n)\nonumber\\
&&+{\mbox{$\cal K$}}({\mbox{\boldmath $k$}},-{\mbox{\boldmath $q$}}
,ik_m,i\Omega_n,-(i\omega_n+i\Omega_n))\Bigr]\:,
\label{Deltaimp}
\end{eqnarray}
where
\begin{eqnarray}\label{calH}
{\mbox{$\cal K$}}&&({\mbox{\boldmath $k$}},{\mbox{\boldmath
$q$}},ik_m,i\Omega_n,i\omega_n)=
{\mbox{$\cal G$}}({\mbox{\boldmath $k$}},ik_m)
\gamma({\mbox{\boldmath $k$}},{\mbox{\boldmath $k$}};ik_m,ik_m+i\Omega_n)
{\mbox{$\cal G$}}({\mbox{\boldmath $k$}},ik_m+i\Omega_n)
\nonumber\\
&&\times
\Gamma({\mbox{\boldmath $k$}},{\mbox{\boldmath $k$}}+{\mbox{\boldmath
$q$}};ik_m+i\Omega_n,ik_m+i\omega_n+i\Omega_n)
{\mbox{$\cal G$}}({\mbox{\boldmath $k$}}+{\mbox{\boldmath
$q$}},ik_m+i\Omega_n+i\omega_n)
\nonumber\\
&&\times
\Gamma({\mbox{\boldmath $k$}}+{\mbox{\boldmath $q$}},{\mbox{\boldmath
$k$}};ik_m+i\omega_n+i\Omega_n,ik_m)\;.
\end{eqnarray}
The factor 2 in Eq. (\ref{Deltaimp}) comes from the spin sum.
In Eq.\ (\ref{calH}),
$k^\alpha\gamma/m$ is the current (vector) vertex function and $\Gamma$
is the charge (scalar) vertex function.  In labeling the variables
in the vertex functions, we use the convention that the incoming
momentum (frequency) is first variable, and the outgoing
momentum  (frequency) is the second variable.
The second term in
the square brackets corresponds to reversing the order
of the two $U_{12}$ lines; see Fig.\ \ref{fig:vertex}.
Eqs.(\ref{Deltaimp})--(\ref{calH}) need to be analytically
continued to the real axis, after which they can be
used as a starting point for evaluating the transconductance in the
weak and strong scattering limits, respectively. It should
be noted that Eq.\ (\ref{calH}) does not include all possible diagrams.
An example of a diagram not included is shown in Fig.\ \ref{fig:crossed}(c).

\subsection{Analytic continuation}

The summation over the fermion frequencies $ik_m$ follows the
standard prescription:\cite{mahan} the discrete sum is
replaced by a contour integration, $\beta^{-1}\sum_{ik_n} f(ik_n) =
-(2\pi i)^{-1}\oint dz n_{\mathrm F}(z) f(z)$.  To evaluate the contour
integral, one must pay attention to the branch cuts of the
integrand, and in the case of Eq.(\ref{Deltaimp}) these
occur at $\mathrm{Im}[z]=0,-\Omega_m,-\Omega_m-\omega_m$ (first term),
and at $\mathrm{Im}[z]=0, -\Omega_m,\omega_m$ (second term).  After
performing the contour integration, we must further set
$i\Omega_m+i\omega_m\to \Omega+\omega + i\delta$,
$i\Omega_m\to \Omega+ i\delta$, and $i\omega_m\to\omega-i\delta$
(see Eq.(\ref{sigma21})).  After some tedious, but
straightforward, algebra one finds
\begin{eqnarray}\label{anal}
\Delta^\alpha({\mbox{\boldmath $q$}},  \Omega+\omega+i\delta,&&\omega-i\delta)=
{2\over \nu m}\sum_{\mbox{\boldmath $k$}}\int_{-\infty}^{\infty}
{d\epsilon\over 2\pi i}n_{\mathrm F}(\epsilon)k^\alpha \nonumber\\
&&\times\Bigl\{ K({\mbox{\boldmath $k$}},{\mbox{\boldmath $q$}},\epsilon,
\Omega,\omega)+K({\mbox{\boldmath $k$}},-{\mbox{\boldmath $q$}},\epsilon,
\Omega,-(\omega+\Omega))\Bigr\}\;,
\end{eqnarray}
where
\begin{eqnarray}\label{H}
K({\mbox{\boldmath $k$}}, && {\mbox{\boldmath
$q$}},\epsilon,\Omega,\omega)=\nonumber\\
&&G^r({\mbox{\boldmath $k$}},\epsilon+\Omega)
\Gamma_{++}({\mbox{\boldmath $k$}},{\mbox{\boldmath $k$}}+{\mbox{\boldmath
$q$}},
\epsilon+\Omega,\epsilon+\omega+\Omega)
G^r({\mbox{\boldmath $k$}}+{\mbox{\boldmath
$q$}},\epsilon+\omega+\Omega)\nonumber\\
&&\times\Bigl\{\Gamma_{++}({\mbox{\boldmath $k$}}+{\mbox{\boldmath
$q$}},{\mbox{\boldmath $k$}},
\epsilon+\omega+\Omega,\epsilon)G^r({\mbox{\boldmath $k$}},\epsilon)
\gamma_{++}({\mbox{\boldmath $k$}},{\mbox{\boldmath $k$}};
\epsilon,\epsilon+\Omega)\nonumber\\
&&-\Gamma_{+-}({\mbox{\boldmath $k$}}+{\mbox{\boldmath $q$}},{\mbox{\boldmath
$k$}},
\epsilon+\omega+\Omega,\epsilon)
G^a({\mbox{\boldmath $k$}},\epsilon)
\gamma_{-+}({\mbox{\boldmath $k$}},{\mbox{\boldmath $k$}};
\epsilon,\epsilon+\Omega)\Bigr\}\nonumber\\
&&+\nonumber\\
&&G^a({\mbox{\boldmath $k$}}+{\mbox{\boldmath $q$}},\epsilon+\omega)
\Gamma_{--}({\mbox{\boldmath $k$}}+{\mbox{\boldmath $q$}},{\mbox{\boldmath
$k$}},
\epsilon+\omega,\epsilon-\Omega)
G^a({\mbox{\boldmath $k$}},\epsilon-\Omega)\nonumber\\
&&\times\Bigl\{\gamma_{-+}({\mbox{\boldmath $k$}},{\mbox{\boldmath $k$}};
\epsilon-\Omega,\epsilon)G^r({\mbox{\boldmath $k$}},\epsilon)
\Gamma_{+-}({\mbox{\boldmath $k$}},{\mbox{\boldmath $k$}}+{\mbox{\boldmath
$q$}},
\epsilon,\epsilon+\omega)\nonumber\\
&&-\gamma_{--}({\mbox{\boldmath $k$}},{\mbox{\boldmath $k$}};
\epsilon-\Omega,\epsilon)
G^a({\mbox{\boldmath $k$}},\epsilon)
\Gamma_{--}({\mbox{\boldmath $k$}},{\mbox{\boldmath $k$}}+{\mbox{\boldmath
$q$}},
\epsilon,\epsilon+\omega)\Bigr\}\nonumber\\
&&+\nonumber\\
&&G^a({\mbox{\boldmath $k$}},\epsilon-\Omega-\omega)
\gamma_{-+}({\mbox{\boldmath $k$}},{\mbox{\boldmath $k$}};
\epsilon-\Omega-\omega,\epsilon-\omega)
G^r({\mbox{\boldmath $k$}},\epsilon-\omega)\nonumber\\
&&\times\Bigl\{\Gamma_{++}({\mbox{\boldmath $k$}},{\mbox{\boldmath
$k$}}+{\mbox{\boldmath $q$}},
\epsilon-\omega,\epsilon)G^r({\mbox{\boldmath $k$}}+{\mbox{\boldmath
$q$}},\epsilon)
\Gamma_{+-}({\mbox{\boldmath $k$}}+{\mbox{\boldmath $q$}},{\mbox{\boldmath
$k$}},
\epsilon,\epsilon-\Omega-\omega)\nonumber\\
&&-\Gamma_{+-}({\mbox{\boldmath $k$}},{\mbox{\boldmath $k$}}+{\mbox{\boldmath
$q$}},
\epsilon-\omega,\epsilon)
G^a({\mbox{\boldmath $k$}}+{\mbox{\boldmath $q$}},\epsilon)
\Gamma_{--}({\mbox{\boldmath $k$}}+{\mbox{\boldmath $q$}},{\mbox{\boldmath
$k$}},
\epsilon,\epsilon-\Omega-\omega)\Bigr\}\;.
\end{eqnarray}
Here the subscripts $\pm\pm$ indicate the signs of the
(infinitesimal) imaginary
parts of the vertex functions' frequency arguments.
This result is still quite general, and  can be
evaluated within different levels  of approximation,
of which we shall illustrate three special cases.

\subsection{The Boltzmann limit ($\omega\tau > 1$ and/or $Dq^2\tau > 1$)}
\label{sec:boltzmann}

In the weak scattering limit, we can neglect the charge
vertex corrections, since $\Gamma$  differs from unity only in
a small region, where $\omega\tau$ and
$D q^2\tau$ are small. Here $\tau$ is the
life-time due to impurity scattering and $D$ is the
diffusion constant.
In the dc and weak-scattering (or ``Boltzmann'') limit,
$\Delta_{\mathrm{B}}$  becomes
\begin{eqnarray}
\Delta^{\alpha}_{\mathrm{B}}
({\mbox{\boldmath $q$}},{\mbox{\boldmath $q$}};\omega+i\delta,&&\omega-i\delta)
=\frac{2}{m\nu}\sum_{\mbox{\boldmath $k$}} \int_{-\infty}^\infty
\frac{d\epsilon}{2\pi i}\,n_{\mathrm{F}}(\epsilon) k^\alpha\nonumber\\
&&\times\Bigl\{
K_{\mathrm{B}}({\mbox{\boldmath $k$}},{\mbox{\boldmath $q$}},\epsilon,
\Omega=0,\omega)+K_{\mathrm{B}}({\mbox{\boldmath $k$}},-{\mbox{\boldmath
$q$}},\epsilon,
\Omega=0,-\omega)\Bigr\}\;,
\label{DeltaR}
\end{eqnarray}
where
\begin{eqnarray}
K_{\mathrm{B}}&&({\mbox{\boldmath $k$}},
{\mbox{\boldmath $q$}},\epsilon,\Omega=0,\omega)=
\nonumber\\
&&
G^r({\mbox{\boldmath $k$}},\epsilon)
G^r({\mbox{\boldmath $k$}}+{\mbox{\boldmath $q$}},\epsilon+\omega)
[G^r({\mbox{\boldmath $k$}},\epsilon)
\gamma_{++}^{\mathrm{B}}
(\mbox{\boldmath $k$},\mbox{\boldmath $k$};\epsilon,\epsilon)
-G^a({\mbox{\boldmath $k$}},\epsilon)
\gamma_{-+}^{\mathrm{B}}
(\mbox{\boldmath $k$},\mbox{\boldmath $k$};\epsilon,\epsilon)
\nonumber \\
+&&G^a({\mbox{\boldmath $k$}}+{\mbox{\boldmath $q$}},\epsilon+\omega)
G^a({\mbox{\boldmath $k$}},\epsilon)
[
\gamma_{-+}^{\mathrm{B}}
(\mbox{\boldmath $k$},\mbox{\boldmath $k$};\epsilon,\epsilon)
G^r({\mbox{\boldmath $k$}},\epsilon)
-
\gamma_{--}^{\mathrm{B}}
(\mbox{\boldmath $k$},\mbox{\boldmath $k$};\epsilon,\epsilon)
G^a({\mbox{\boldmath $k$}},\epsilon)]
\nonumber\\
+&&G^a({\mbox{\boldmath $k$}},\epsilon-\omega)\gamma_{-+}^{\mathrm{B}}
(\mbox{\boldmath $k$},\mbox{\boldmath
$k$};\epsilon-\omega,\epsilon-\omega)
G^r({\mbox{\boldmath $k$}},\epsilon-\omega)
[G^r({\mbox{\boldmath $k$}}+{\mbox{\boldmath $q$}},\epsilon)-
G^a({\mbox{\boldmath $k$}}+{\mbox{\boldmath $q$}},\epsilon)]\nonumber\\
\approx&&\gamma_{-+}^{\mathrm{B}}(\mbox{\boldmath $k$},\mbox{\boldmath
$k$};\epsilon_{\mbox{\boldmath $k$}},\epsilon_{\mbox{\boldmath $k$}})
\Bigl\{-G^r({\mbox{\boldmath $k$}},\epsilon)
G^a({\mbox{\boldmath $k$}},\epsilon)
[G^r({\mbox{\boldmath $k$}}+{\mbox{\boldmath $q$}},\epsilon+\omega)
-G^a({\mbox{\boldmath $k$}}+{\mbox{\boldmath $q$}},\epsilon+\omega)]\nonumber\\
+&&G^r({\mbox{\boldmath $k$}},\epsilon-\omega)G^a({\mbox{\boldmath
$k$}},\epsilon-\omega)
[G^r({\mbox{\boldmath $k$}}+{\mbox{\boldmath $q$}},\epsilon)-
G^a({\mbox{\boldmath $k$}}+{\mbox{\boldmath $q$}},\epsilon)]\Bigr\}
\;.
\label{HB}
\end{eqnarray}
In writing the approximate equation, we used the fact
that in the Boltzmann limit,
(i) the terms of the form $G^rG^r$ or $G^aG^a$ (with
equal momentum and frequency arguments) are
smaller by a factor of $1/\tau E_{F} \ll 1$ than terms of the
type $G^rG^a$, and can hence be neglected
in the present level of approximation, and (ii)
$G^r(\mbox{\boldmath $k$},\omega)G^a(\mbox{\boldmath $k$},\omega)
=\tau A(\mbox{\boldmath $k$},\omega)$, where $A(\mbox{\boldmath
$k$},\omega)$ is the spectral function
and $1/2\tau=-{\mathrm{Im}}\Sigma^r(\mbox{\boldmath $k$},
\epsilon_{\mbox{\boldmath $k$}})$.
$A(\mbox{\boldmath $k$},\omega)$ is sharply peaked around
$\epsilon = \epsilon_{\mbox{\boldmath $k$}}$, and hence the
energy arguments in $\gamma_{-+}$ can be replaced
by $\epsilon_{\mbox{\boldmath $k$}}$.

In the Boltzmann limit of the current vertex function $\gamma^{\mathrm{B}}$
is well-known:\cite{mahan7.1.c}
\begin{equation}
\gamma_{-+}^{\mathrm{B}}({\mbox{\boldmath $k$}},{\mbox{\boldmath $k$}};
\epsilon_{\mbox{\boldmath $k$}},\epsilon_{\mbox{\boldmath $k$}})
= \frac{\tau_{\mathrm{tr}}({\mbox{\boldmath $k$}})}
{\tau({\mbox{\boldmath $k$}})}.
\label{gammaboltz}
\end{equation}
Using $G^r(\mbox{\boldmath $k$},\omega)-G^a(\mbox{\boldmath
$k$},\omega) =-iA(\mbox{\boldmath $k$},\omega)$,
$G^r(\mbox{\boldmath $k$},\omega)G^a(\mbox{\boldmath $k$},\omega)
=\tau A(\mbox{\boldmath $k$},\omega)$ and Eq.\ (\ref{gammaboltz}),
Eq.(\ref{DeltaR}) thus simplifies to
\begin{eqnarray}
\Delta^\alpha_{\mathrm{B}}({\mbox{\boldmath $q$}},{\mbox{\boldmath
$q$}};\omega+i\delta
,\omega-i\delta)
&\approx&\frac{2}{m\nu}\sum_{\mbox{\boldmath $k$}}
\int_{-\infty}^\infty
\frac{d\epsilon}{2\pi}\,n_{\mathrm{F}}(\epsilon) k^{\alpha}
\tau_{\mathrm{tr}}({\mbox{\boldmath $k$}})
\nonumber \\&&
\times\left[
A({\mbox{\boldmath $k$}},\epsilon)\left\{A({\mbox{\boldmath
$k$}}-{\mbox{\boldmath $q$}},\epsilon-\omega)+
A({\mbox{\boldmath $k$}}+{\mbox{\boldmath $q$}},\epsilon+\omega)
\right\}
\right. \nonumber \\&&\left.
-A({\mbox{\boldmath $k$}},\epsilon+\omega)A({\mbox{\boldmath
$k$}}-{\mbox{\boldmath $q$}},\epsilon)-
A({\mbox{\boldmath $k$}},\epsilon-\omega)A({\mbox{\boldmath
$k$}}+{\mbox{\boldmath $q$}},\epsilon)
\right] \nonumber \\&=&
{2\over m\nu}\sum_{\mbox{\boldmath $k$}}\int_{-\infty}^\infty{d\epsilon\over
2\pi}
[n_{\mathrm{F}}(\epsilon+\omega)-n_{\mathrm{F}}(\epsilon)]\nonumber \\&&
\times\bigl[
({\mbox{\boldmath $k$}}+{\mbox{\boldmath $q$}})^\alpha
\tau_{\mathrm{tr}}({\mbox{\boldmath $k$}}+{\mbox{\boldmath $q$}})
-{\mbox{\boldmath $k$}}^\alpha\tau_{\mathrm{tr}}({\mbox{\boldmath $k$}})\bigr]
A({\mbox{\boldmath $k$}}+{\mbox{\boldmath $q$}},\epsilon+\omega)
A({\mbox{\boldmath $k$}},\epsilon)\;.
\label{Deltapm}
\end{eqnarray}
The Boltzmann limit is recovered by using free Green
functions, which implies that  the spectral functions reduce
to $\delta$-functions. We find
\begin{equation}\label{Deltafinal}
\Delta^\alpha_{\mathrm{B}}({\mbox{\boldmath $q$}},{\mbox{\boldmath $q$}}
;\omega+i\delta,\omega-i\delta)=
\frac{2\bar{\tau}_{\mathrm{tr}}}{m}\,
F^\alpha({\mbox{\boldmath $q$}},\omega),
\end{equation}
where the {\em transport polarization} $F^\alpha$ is given by
\begin{equation}
F^\alpha({\mbox{\boldmath $q$}},\omega) = \frac{2}{\nu\bar{\tau}_{\mathrm{tr}}}
{\mathrm{Im}}
\sum_{\mbox{\boldmath $k$}}\frac{n_{\mathrm{F}}
(\epsilon_{{\mbox{\boldmath $k$}}+{\mbox{\boldmath $q$}}})-
n_{\mathrm{F}}(\epsilon_{\mbox{\boldmath $k$}})}
{\epsilon_{{\mbox{\boldmath $k$}}+{\mbox{\boldmath
$q$}}}-\epsilon_{{\mbox{\boldmath $k$}}}-
\omega-i\delta}\left[
({\mbox{\boldmath $k$}}+{\mbox{\boldmath $q$}})^\alpha
\tau_{\mathrm{tr}}({\mbox{\boldmath $k$}}+{\mbox{\boldmath $q$}})
-{\mbox{\boldmath $k$}}^\alpha\tau_{\mathrm{tr}}({\mbox{\boldmath
$k$}})\right].
\end{equation}
Here $\bar{\tau}_{\mathrm{tr}}$ determines the in-plane
conductivity,
$\sigma_{ii} = e^2 n_i \bar{\tau}_{\mathrm{tr},i}/m_i.$
When (\ref{Deltafinal}) is inserted into the expression for
transconductivity, Eq.(\ref{sigma21}), we obtain
\begin{equation}\label{rho21boltzmann}
\rho_{21} = -\frac{\sigma_{21}}{\sigma_{11}\sigma_{22}}
=-\frac{1}{2 n_1n_2}\frac{1}{\nu}\sum_{\mbox{\boldmath $q$}}
\int_{-\infty}^\infty \frac{d\omega}{2\pi}\frac{\beta}{\sinh^2[\beta\omega/2]}
\left|U_{12}(q,\omega)\right|^2
F_1({\mbox{\boldmath $q$}},\omega) F_2({\mbox{\boldmath $q$}},\omega),
\label{rho21}
\end{equation}

Several comments are now in order. Without an applied magnetic
field, the transconductivity and consequently also the transresistivity
are diagonal in the Cartesian coordinates, and we have suppressed
the $\{\alpha\beta\}$ indices in (\ref{rho21boltzmann}).
For constant $\tau$'s
transport polarization is related to the (bare) RPA polarization
function,
$F^\alpha({\mbox{\boldmath $q$}},\omega) =  q^\alpha{\mathrm{Im}}\,
\chi_0({\mbox{\boldmath $q$}},\omega)$.  In this limit
Eq.\ (\ref{rho21boltzmann})
reproduces the standard result for transresistivity,
see, {\em e.g.}  Refs. \onlinecite{jauh93,zhen93}.
Since the above derivation ignores all higher order
and/or quantum mechanical processes, it is not surprising that
one can
derive (\ref{rho21boltzmann})
directly from the Boltzmann equation.\cite{KarstenBenlong}
We also emphasize that in general the drag rate, or the
transresistivity, cannot be expressed in terms of the polarization
function; rather, one must use the more general object $F^\alpha$
defined above.

\subsection{Diffusive limit ($\omega\tau < 1$, $Dq^2\tau < 1$)}
\label{sec:ladder}

In this section we evaluate the transconductivity,
in the weak scattering limit and in the diffusive limit
($D q^2\tau < 1$ and $\omega\tau < 1$), including vertex corrections.
Specifically, we consider momentum independent
relaxation times, in which case $\tau_{\mathrm{tr}}=\tau$, and include
vertex corrections due to ladder diagrams.
Then, we have $\gamma\equiv 1$ for all $\pm$ combinations, and
the charge vertex is given by\cite{alts79}
\begin{equation}\label{Gammaladder}
\Gamma({\mbox{\boldmath $k$}},{\mbox{\boldmath $k$}}+{\mbox{\boldmath
$q$}},z_1,z_2) =
{\theta[-{\mathrm{Im}}(z_1){\mathrm{Im}}(z_2)]
\over
\tau\{Dq^2-i(z_1-z_2){\mathrm{sgn[Im}}(z_1-z_2)]\}}
+\theta[{\mathrm{Im}}(z_1){\mathrm{Im}}(z_2)]\;,
\end{equation}
where $\theta$ is the step function.
It is now straightforward to use (\ref{Gammaladder}) in the
$K$-function (\ref{H}).  Including only terms which involve
$G^rG^a$ (with same arguments), and introducing a short-hand
notation
\begin{equation}\label{Gammapm}
\Gamma^{\pm}(q,\omega) = [\tau(Dq^2\pm i\omega)]^{-1}\;,
\end{equation}
allows us to write
\begin{eqnarray}\label{HL+}
&&K_{\mathrm{L}}({\mbox{\boldmath $k$}},\pm{\mbox{\boldmath
$q$}},\epsilon,\Omega=0,\pm\omega)=
G^r({\mbox{\boldmath $k$}},\epsilon)G^a({\mbox{\boldmath
$k$}},\epsilon)\nonumber\\
&&\times
\Bigl[
-G^r({\mbox{\boldmath $k$}}\pm{\mbox{\boldmath
$q$}},\epsilon\pm\omega)\Gamma^{\mp}(q,\omega)
+G^a({\mbox{\boldmath $k$}}\pm{\mbox{\boldmath
$q$}},\epsilon\pm\omega)\Gamma^{\pm}(q,\omega)
\Bigr]
\nonumber\\
&&+
G^r({\mbox{\boldmath $k$}},\epsilon\mp\omega)G^a({\mbox{\boldmath
$k$}},\epsilon\mp\omega)
\Bigl[
G^r({\mbox{\boldmath $k$}}\pm{\mbox{\boldmath
$q$}},\epsilon)\Gamma^{\mp}(q,\omega)
-G^a({\mbox{\boldmath $k$}}\pm{\mbox{\boldmath
$q$}},\epsilon)\Gamma^{\pm}(q,\omega)
\Bigr]\:.
\end{eqnarray}
The triangle function in the ladder approximation,
$\Delta_{\mathrm{L}}^{\alpha}$, then
becomes
\begin{eqnarray}\label{Deltaladder}
\Delta_{\mathrm{L}}^{\alpha}&&({\mbox{\boldmath $q$}},{\mbox{\boldmath
$q$}},\omega+i\delta,
\omega-i\delta) =
{2\tau\over m\nu}\sum_{\mbox{\boldmath $k$}}{\mbox{\boldmath $k$}}^{\alpha}
\int_{-\infty}^{\infty} {d\epsilon\over 2\pi}
[n_{\mathrm{F}}(\epsilon+\omega)-n_{\mathrm{F}}(\epsilon)]\nonumber\\
&&\times\Bigl\{
2{\mathrm{Im}}\bigl[\Gamma^-(q,\omega)
G^r({\mbox{\boldmath $k$}}+{\mbox{\boldmath
$q$}},\epsilon+\omega)\bigr]A({\mbox{\boldmath $k$}},\epsilon)
-
2{\mathrm{Im}}\bigl[\Gamma^+(q,\omega)
G^r({\mbox{\boldmath $k$}}-{\mbox{\boldmath
$q$}},\epsilon)\bigr]A({\mbox{\boldmath $k$}},\epsilon+\omega)
\Bigr\}
\;.
\end{eqnarray}
We observe that the constant-$\tau$ Boltzmann result is readily recovered
from (\ref{Deltaladder}) by replacing $\Gamma$'s by unity.
A generalization to energy-dependent scattering rates
is straightforward, but we do not reproduce the
cumbersome results here.

The next task is to establish a connection between the dressed
polarization function
$\chi({\mbox{\boldmath $q$}},\omega)$ and the triangle function
$\Delta_{\mathrm{L}}$.
In Appendix C we show that
\begin{equation}\label{ImX}
{\mathrm{Im}}\chi({\mbox{\boldmath $q$}},\omega)=
-{2\over\nu}\sum_{\mbox{\boldmath $k$}}
{\mathrm{Im}}\Bigl\{
\int{d\epsilon\over 2\pi i}
[n_{\mathrm{F}}(\epsilon+\omega)-n_{\mathrm{F}}(\epsilon)]
\Gamma^-(q,\omega)G^r({\mbox{\boldmath $k$}}+{\mbox{\boldmath
$q$}},\epsilon+\omega)
G^a({\mbox{\boldmath $k$}},\epsilon)\Bigr\}\;.
\end{equation}
Comparison of (\ref{Deltaladder}) and (\ref{ImX}) reveals some
similarity, but clearly few more steps are required.  We complete
the connection by making a few observations.  First, we express
the spectral functions in (\ref{Deltaladder}) in terms of
the retarded and advanced functions.  The resulting integrals can
be grouped in two classes: (i) integrals involving products of the
type $G^rG^a$, and (ii) integrals involving products of the form
$G^rG^r$ or $G^aG^a$.  We have earlier argued that type-(ii) integrals
can be neglected in comparison with type-(i) integrals, if the
momentum and frequency variables are equal.  We now state that,
in the present weak-scattering limit,
this criterion applies also to functions with momentum variables
and frequency variables which differ by less than ${D\tau}^{-1/2}$
and $\tau^{-1}$ (the diffusive limit).  A proof for this statement
is given in Appendix \ref{app:proof}.
Thus, keeping only the $G^rG^a$-terms in (\ref{Deltaladder}) allows
us to express the quantity in curly brackets as
\begin{eqnarray}
\{ \cdots \} \to && -\Bigl[\Gamma^-(q,\omega)
G^r({\mbox{\boldmath $k$}}+{\mbox{\boldmath $q$}},\epsilon+\omega)
G^a({\mbox{\boldmath $k$}},\epsilon)+ c.c \Bigr]\nonumber\\
&& +
\Bigl[\Gamma^-(q,\omega)G^r({\mbox{\boldmath $k$}},\epsilon+\omega)
G^a({\mbox{\boldmath $k$}}-{\mbox{\boldmath $q$}},\epsilon)+ c.c
\Bigr]\;.\nonumber
\end{eqnarray}
In the above analysis the second term can be made to coincide
with the first one by shifting the summation variable
${\mbox{\boldmath $k$}}\to{\mbox{\boldmath $k$}}+{\mbox{\boldmath $q$}}$;
however when
doing this the prefactor ${\mbox{\boldmath $k$}}^{\alpha}$ in
(\ref{Deltaladder})
generates
an extra ${\mbox{\boldmath $q$}}^{\alpha}$.  This is exactly what
is needed to give the required result,
\begin{equation}
\Delta^{\alpha}_{\mathrm{L}}({\mbox{\boldmath $q$}},{\mbox{\boldmath
$q$}},\omega+i\delta,
\omega-i\delta)
= {2\tau{\mbox{\boldmath $q$}}^{\alpha}\over m}
{\mathrm{Im}}\chi({\mbox{\boldmath $q$}},\omega)\;.
\label{deleqchi}
\end{equation}

The above analysis shows the equivalence of
the triangle function and the polarization
function  in the small $q$ and $\omega$ limit, confirming
the result obtained by Zheng and MacDonald\cite{zhen93} with a
different method.
As observed by these authors, in the high-mobility
samples studied so far the replacement
$\chi_0\to \chi$ does not appear to be important; however, in dirtier samples
the consequences of the vertex corrections
({\it i.e.} full $\chi$) may well become detectable.

\subsection{Weak localization correction to Coulomb drag}
\label{sec:WL}
In the previous sections we included the leading
order impurity scattering diagrams which gave us
the Boltzmann equation result for the case of weak scattering,
and showed how the bare polarization function in a certain
parameter range must be replaced by the dressed polarization
function.
Here we develop the analysis further and
calculate the quantum correction associated with
weak localization (the basic physics of weak localization
is reviewed, {\it e.g.} by Lee and Ramakrishnan\cite{lee85}).
The corrections will be of
the order of $1/(k_F\ell)\ll 1$, where $\ell=v_F\tau$ is the
elastic mean free path.

In Fig.\ \ref{fig:crossed} we display the different types
of crossed diagrams that exist for the function $\Delta$.
The maximally crossed one is the one shown in Fig.\ \ref{fig:crossed}(c).
This diagram is, however,
smaller than the one showed in Fig.\ \ref{fig:crossed}(a), because
of the restricted phase space. The two Green functions attached
to the current vertex in the diagram in Fig.\ \ref{fig:crossed}(a) have
the same arguments because in the limit
$({\mbox{\boldmath $Q$}}=0,\Omega=0)$ the current vertex
leaves the momenta and energies of the entering and leaving
Green functions
unchanged. Therefore there is the possibility
of two overlapping spectral functions giving a overall factor
of $\tau$. This does not happen for the diagram in Fig.\ \ref{fig:crossed}(c).
Neither does the diagram in Fig.\ \ref{fig:crossed}(b)
lead to overlapping spectral functions except
in all very small region of $q,\omega$ space, where $q$ and $\omega$
are the incoming  quantities at the charge vertices. Since we are
integrating over $q,\omega$ the (logarithmic) singularity caused
by the maximally crossed diagrams becomes regularized.
In other words, the contribution from this diagram is small
for the same reasons that the dressing of the charge vertexes,
discussed in Sec.\ \ref{sec:ladder} and can be
neglected for experimentally relevant parameters.\cite{zhen93}.
We therefore conclude that diagrams of the Fig.\ \ref{fig:crossed}(a)-type
dominate the quantum correction to the drag rate.

The leading quantum correction
is given as the sum of the maximally crossed diagrams,
the Cooperon. The resulting vertex
function describing the weak localization correction,
$\gamma^{\mathrm{WL}}$, obeys
\begin{equation}\label{gammawl}
\gamma^{\mathrm{WL}}({\mbox{\boldmath $k$}},{\mbox{\boldmath
$k$}};ik_m,ik_m+i\Omega_n)
= \frac{1}{\nu}\sum_{{\mbox{\boldmath $k$}}'}\frac{{\mbox{\boldmath
$k$}}\cdot{\mbox{\boldmath $k$}}'}{(k')^2}
{\mbox{$\cal G$}}({\mbox{\boldmath $k$}}',ik_m){\mbox{$\cal
G$}}({\mbox{\boldmath $k$}}',ik_m+i\Omega_n)
{\mbox{$\cal C$}}({\mbox{\boldmath $k$}},{\mbox{\boldmath
$k$}}';ik_m,ik_m+i\Omega_n),
\end{equation}
where the Cooperon is given by\cite{ramm86}
\begin{equation}\label{cooperon}
{\mbox{$\cal C$}}({\mbox{\boldmath $k$}},{\mbox{\boldmath
$k$}};ik_m,ik_m+i\Omega_n)=
\frac{1}{2\pi\rho\tau}\frac{1}{1-\zeta(k;ik_m,ik_m+i\Omega_n)},
\end{equation}
where
\begin{equation}
\zeta({\mbox{\boldmath $k$}},{\mbox{\boldmath $k$}}';ik_m,ik_m+i\Omega_n)=
\frac{1}{2\pi\rho\tau} \frac{1}{\nu}\sum_{{\mbox{\boldmath $p$}}}
{\mbox{$\cal G$}}({\mbox{\boldmath $p$}}+{\mbox{\boldmath
$Q$}},ik_m+i\Omega_n){\mbox{$\cal G$}}({\mbox{\boldmath $p$}},ik_m),
\label{zeta}
\end{equation}
where ${\mbox{\boldmath $Q$}}={\mbox{\boldmath $k$}}+{\mbox{\boldmath $k$}}'$.

In order to evaluate the $\zeta$-function, we make use
of the fact that the weak localization divergence
occurs for small $Q$. With this in mind we replace
$\varepsilon({\mbox{\boldmath $k$}}+{\mbox{\boldmath $Q$}})$ by
$\varepsilon({\mbox{\boldmath $k$}})+
v_FQ\cos(\theta)$.
Then we can integrate over ${\mbox{\boldmath $k$}}'$ in Eq.\ (\ref{zeta}).
For small $DQ^2$ and $\Omega\tau$ we obtain
\begin{equation}
\zeta({\mbox{\boldmath $k$}},{\mbox{\boldmath $k$}}';ik_m,ik_m+i\Omega_n)=
\left\{
\begin{array}{cc}
0&,\,\mbox{if}\ (k_m+\Omega_n)k_m<0\\
1-|\Omega_n|\tau-DQ^2\tau&,\,
\mbox{if}\ (k_m+\Omega_n)k_m>0
\end{array}\right. ,
\end{equation}
where the diffusion constant is defined as $D=v_F^2\tau/2$.
This expression is only valid for $Q\ell < 1$,
therefore the upper limit of the integral in Eq.\ (\ref{zeta})
has to be cut off by $1/\ell$.

Next we perform the analytic continuations that are
needed for the evaluation of the function $\Delta$.
Since the analytic continuation $\omega_n\rightarrow\omega\pm i\delta$
leads to $|\omega_n|\rightarrow \mp i\omega$, we obtain for
the $\zeta$-function
\begin{eqnarray}
&&\zeta_{++}=\zeta_{--}=0,\nonumber\\
&&\zeta_{-+}({\mbox{\boldmath $k$}},{\mbox{\boldmath
$k$}}';\epsilon,\epsilon+\Omega)
=\zeta_{-+}({\mbox{\boldmath $k$}},{\mbox{\boldmath
$k$}}';\epsilon-\Omega,\epsilon)\\
&&=1+i\Omega\tau-DQ^2\tau.\nonumber
\end{eqnarray}
The Cooperon that enters
the expression for $K_{WL}$, Eq.(\ref{H}), then acquires the familiar
form
\begin{equation}
C_{-+} = \frac{1}{2\pi\rho\tau^2}\frac{1}{-i\Omega +DQ^2}.
\end{equation}
After integration over ${\mbox{\boldmath $Q$}}$ we find the
weak localization vertex function,
\begin{eqnarray}
&&\gamma^{\mathrm{WL}}({\mbox{\boldmath $k$}},{\mbox{\boldmath $k$}};
\epsilon-\Omega,\epsilon) =
\gamma^{\mathrm{WL}}({\mbox{\boldmath $k$}},{\mbox{\boldmath $k$}};
\epsilon,\epsilon+\Omega)\nonumber\\
&&\approx A(k,\epsilon) \frac{1}{\pi k_F\ell\tau}\ln(\Omega\tau)
\equiv A(k,\epsilon) \eta^{\mathrm{WL}}(\Omega)/2\tau.
\end{eqnarray}
Here $\eta^{\mathrm{WL}}$ is
the ratio between the quantum correction and the classical
conductivity:
$\eta^{\mathrm{WL}}(\Omega)=\delta \sigma(\omega)/\sigma_0$.\cite{ramm86}
The combinations $\gamma^{\mathrm{WL}}_{++}$ and $\gamma^{\mathrm{WL}}_{--}$
are both zero.
We now get for the $K$-function
\begin{equation}
K_{\mathrm{WL}}({\mbox{\boldmath $k$}},{\mbox{\boldmath
$q$}};\epsilon,\Omega,\omega)
\approx -i\eta^{\mathrm{WL}} \Bigl\{-[A({\mbox{\boldmath $k$}},\epsilon)]^2
A({\mbox{\boldmath $k$}}+{\mbox{\boldmath $q$}},\epsilon+\omega)
+ [A({\mbox{\boldmath $k$}},\epsilon-\omega)]^2 A({\mbox{\boldmath
$k$}}+{\mbox{\boldmath $q$}},\epsilon)
\Bigr\}
\end{equation}
Using that $A^2 \approx \tau A/2$ for large $\tau$ we can
express the weak localization correction
$\delta\Delta_{\mathrm{WL}}$ in terms of
the response functions as
\begin{eqnarray}
\delta\Delta^\alpha_{\mathrm{WL}}({\mbox{\boldmath $q$}},{\mbox{\boldmath
$q$}};\omega+i\delta,
\omega-i\delta,\Omega)=
\frac{2\tau\eta^{\mathrm{WL}}(\Omega)}{m}
\, q^\alpha \mathrm{Im}\, \chi_0({\mbox{\boldmath
$q$}},\omega)=\eta^{\mathrm{WL}}(\Omega)
\Delta^\alpha_{\mathrm{B}}({\mbox{\boldmath $q$}},{\mbox{\boldmath
$q$}};\omega+i\delta,
\omega-i\delta),
\end{eqnarray}
which immediately leads to the conclusion that to leading order
the weak localization correction to transconductance is
\begin{equation}
\delta\sigma_{21}^{\mathrm{WL}}(\Omega)= [\eta^{\mathrm{WL}}_1(\Omega)+
\eta^{\mathrm{WL}}_2(\Omega)]\sigma_{21}^0.
\end{equation}
Consequently, the transresistance is {\em unaffected}
by the weak localization correction because
\begin{equation}
\rho_{21} \approx
-\frac{\sigma_{21}^0(1+\eta_1+\eta_2)}
{\sigma_{11}^0(1+\eta_1)\sigma_{22}^0(1+\eta_2)}
\approx \rho_{21}^0.
\end{equation}
Weak localization is strongly affected by external magnetic
fields.  The formalism presented above can be extended to
include magnetic fields; in particular, the topology of all
diagrams remains unaltered.

\section{Finite frequency response at $T=0$}
\label{sec:zero}

\subsection{General expression}

For finite frequencies and finite temperatures the
analysis becomes considerably more complicated and for
simplicity we therefore restrict ourselves to study the
finite frequencies case at zero temperature.
Furthermore, we will consider a model
where all vertex functions can be replaced by unity,
i.e. a system with short range impurity potentials and
a system not in the diffusive limit. The transconductivity
in terms of the time-ordered current-current
correlation function is
\begin{equation}
\sigma^{\alpha\beta}(\mbox{\boldmath $Q$},\Omega)
=\frac{e^2}{\Omega}\left(
i\mbox{Re}\,\left[\Pi^{\alpha\beta}_t(\mbox{\boldmath $Q$},\Omega)\right]
-\mbox{sgn}(\Omega)\mbox{Im}\,
\left[\Pi^{\alpha\beta}_t(\mbox{\boldmath $Q$},\Omega)\right] \right)
\end{equation}
The time-ordered current-current correlation function
is now written in terms of time-ordered Green functions as
\begin{equation}
\Pi^{\alpha\beta}_t(\mbox{\boldmath $Q$}=0,\Omega) = -{i\over 2}
\int{{d\omega}\over{2\pi}}
\frac{1}{\nu}\sum_{\mbox{\boldmath $q$}}
|U_{12}(q)|^2 \Delta^\alpha_t(\mbox{\boldmath $q$},\mbox{\boldmath
$q$};\Omega+\omega,\omega)
\Delta^\beta_t(-\mbox{\boldmath $q$},-\mbox{\boldmath
$q$};-\Omega-\omega,-\omega),
\label{eq:zerocorr}
\end{equation}
where
\begin{eqnarray}
\Delta^\beta_t(\mbox{\boldmath $q$},\mbox{\boldmath $q$};\Omega+\omega,\omega)
 &=&
\int\frac{d\omega_1}{2\pi}
\frac{2}{\nu}\sum_{\mbox{\boldmath $k$}}
v_{\alpha}(\mbox{\boldmath $k$})
G^t(\mbox{\boldmath $k$},\omega_1 + {1\over 2}\Omega)
G^t({\mbox{\boldmath $k$}},\omega_1 - {1\over 2}\Omega)\nonumber\\
&&\times \left [G^t({\mbox{\boldmath $k$}} + {\mbox{\boldmath $q$}},\omega_1 +
\omega)
+ G^t({\mbox{\boldmath $k$}} - {\mbox{\boldmath $q$}},\omega_1 - \omega)\right
].
\label{eq:zeroDelta}
\end{eqnarray}
It is straightforward to see that
$\Delta_t^\alpha(\mbox{\boldmath $q$},\mbox{\boldmath $q$};\omega,\omega)=0$,
which can be
shown along the same lines as in the last part of
Appendix \ref{app:zero}. Furthermore we can show
that $\Delta(\mbox{\boldmath $q$},\mbox{\boldmath $q$};\omega,\omega+\Omega)$
is proportional to $\Omega$, and consequently the
transconductivity vanishes at zero frequency and zero temperature
in agreement with Eq.\ (\ref{sigma21}).

\subsection{Clean system}
To evaluate the zero-temperature correlation function it is useful
to decompose the time-ordered Green function into advanced and retarded
parts according to $G^c({\mbox{\boldmath $q$}},\omega) = \Theta(\omega)
G^r({\mbox{\boldmath $q$}},\omega) + \Theta(-\omega)G^a({\mbox{\boldmath
$q$}},\omega)$.
In the case of a clean system, the decomposition can also be carried
out in momentum space as $G^c({\mbox{\boldmath $q$}},\omega) = \Theta(|
{\mbox{\boldmath $q$}}| - k_F)G^r({\mbox{\boldmath $q$}},\omega) + \Theta(k_F -
|{\mbox{\boldmath $q$}}|)G^a({\mbox{\boldmath $q$}},\omega)$, which is actually
more
convenient since it leaves the frequency integrals unrestricted.
Consequently, using the momentum space decomposition, we can carry out
the conventional pole-position analysis and find that most of the
terms arising from (\ref{eq:zerocorr}) vanish. The remaining non-zero
terms are most conveniently evaluated using the frequency-space
decomposition, which yields for $\Omega > 0$
\begin{eqnarray}
&&\Pi^{\alpha\alpha}_t({\mbox{\boldmath $Q$}}=0,\Omega) =
 -\frac{4i}{\nu^3}\sum_{{\mbox{\boldmath $q$}},{\mbox{\boldmath
$p$}},{\mbox{\boldmath $k$}}}
|U_{12}(q)|^2
v_\alpha({\mbox{\boldmath $k$}}) v_\alpha({\mbox{\boldmath $p$}})\nonumber\\
&\times&\Bigl\{
2\int_{-\Omega/2}^{\Omega/2}{{d\omega_1}\over{2\pi}}
\int_{-\Omega/2}^{\omega_1}{{d\omega_2}\over{2\pi}}
\int_{\omega_2}^{\omega_1}{{d\omega}\over{2\pi}}
G^r({\mbox{\boldmath $k$}},\omega_1 + {1\over 2}\Omega)
G^a({\mbox{\boldmath $k$}},\omega_1 - {1\over 2}\Omega)\nonumber\\
&\times&
G^r({\mbox{\boldmath $k$}} - {\mbox{\boldmath $q$}},\omega_1 - \omega)
G^r({\mbox{\boldmath $p$}},\omega_2 + {1\over 2}\Omega)
G^a({\mbox{\boldmath $p$}},\omega_2 - {1\over 2}\Omega)
G^a({\mbox{\boldmath $p$}} - {\mbox{\boldmath $q$}},\omega_2 -
\omega)\nonumber\\
&+& \int_{-\Omega/2}^{\Omega/2}{{d\omega_1}\over{2\pi}}
\int_{-\omega_1}^{\Omega/2}{{d\omega_2}\over{2\pi}}
\int_{-\omega_2}^{\omega_1}{{d\omega}\over{2\pi}}
G^r({\mbox{\boldmath $k$}},\omega_1 + {1\over 2}\Omega)
G^a({\mbox{\boldmath $k$}},\omega_1 - {1\over 2}\Omega)\nonumber\\
&\times&
G^r({\mbox{\boldmath $k$}} - {\mbox{\boldmath $q$}},\omega_1 - \omega)
G^r({\mbox{\boldmath $p$}},\omega_2 + {1\over 2}\Omega)
G^a({\mbox{\boldmath $p$}},\omega_2 - {1\over 2}\Omega)
G^r({\mbox{\boldmath $p$}} + {\mbox{\boldmath $q$}},\omega_2 +
\omega)\nonumber\\
&+& \int_{-\Omega/2}^{\Omega/2}{{d\omega_1}\over{2\pi}}
\int_{-\Omega/2}^{-\omega_1}{{d\omega_2}\over{2\pi}}
\int_{\omega_1}^{-\omega_2}{{d\omega}\over{2\pi}}
G^r({\mbox{\boldmath $k$}},\omega_1 + {1\over 2}\Omega)
G^a({\mbox{\boldmath $k$}},\omega_1 - {1\over 2}\Omega)\nonumber\\
&\times&
G^a({\mbox{\boldmath $k$}} - {\mbox{\boldmath $q$}},\omega_1 - \omega)
G^r({\mbox{\boldmath $p$}},\omega_2 + {1\over 2}\Omega)
G^a({\mbox{\boldmath $p$}},\omega_2 - {1\over 2}\Omega)
G^a({\mbox{\boldmath $p$}} + {\mbox{\boldmath $q$}},\omega_2 + \omega)\Bigr\}.
\label{eq:AR}
\end{eqnarray}
Since all frequency integrations run at most over an interval
of length $\Omega$, the end result is proportional to
$\Omega^3$. We can furthermore show that $\Pi_{\alpha\alpha}^c(
{\mbox{\boldmath $Q$}},\Omega) = \Pi_{\alpha\alpha}^c({\mbox{\boldmath
$Q$}},-\Omega)$,
so that $\Pi_{\alpha\alpha}^c({\mbox{\boldmath $Q$}},\Omega) \sim |\Omega|^3$.
Using the fluctuation-dissipation theorem we find
$\text{Re\ }\Pi^r \sim |\Omega|^3$ and $\text{Im\ }\Pi^r \sim \Omega^3$.
Thus, to a leading order in $\tau$,
the real part of the transconductivity is proportional
to $\Omega^2$, and its imaginary part
is proportional to ${\mathrm{sgn}}(\Omega)\Omega^2$.

\subsection{Disordered systems}
For disordered systems we can only use the frequency-space
decomposition, and consequently the pole-position analysis is not
quite as powerful as for clean systems. We can, however, determine
the leading corrections by regarding $\tau^{-1}$ as a perturbation
and using a Taylor expansion of the type
$G^r(\omega_1)G^r(\omega_1)G^r(\omega_1 - \omega) =
G_0^r(\omega_1)G_0^r(\omega_1)G_0^r(\omega_1 - \omega)
+ (i/2\tau)(\partial/\partial\omega_1)[G_0^r(\omega_1)G_0^r(\omega_1)
G_0^r(\omega_1 - \omega)]$.
After the expansion, all propagators are given by clean system
Green functions, and we can easily do the integral over one of the
frequency arguments
(in the example over $\omega_1$). The resulting
term is identical with a term that is encountered in the evaluation
of an auxiliary time-ordered function for a clean system. The auxiliary
function can be analyzed also by means of the momentum-space
decomposition (since all propagators are given by $G_0^c$), and the
frequency dependence of the various terms can be obtained in a manner similar
to what we did in the clean system case. Carrying out the
analysis for all terms arising from  (\ref{eq:zerocorr}), we find that
the leading order corrections to $\Pi^c$ are of the form
$\Omega^2/\tau$, and we obtain
\begin{equation}
\Pi_t({\mbox{\boldmath $Q$}},\Omega) =
F_0({\mbox{\boldmath $Q$}})|\Omega|^3 + F_1({\mbox{\boldmath
$Q$}})\Omega^2/\tau.
\end{equation}
Thus, we finally have
\begin{equation}
\sigma_{21}({\mbox{\boldmath $Q$}} = 0, \Omega) \sim \left\{
\begin{array}{ll}
\Omega,&|\Omega| \ll 1/\tau\\
\Omega^2,&1/\tau\ll|\Omega|\ll\epsilon_F/\hbar
\end{array}
\right .
\end{equation}

The constant $F_0(0)$ can be evaluated approximately be keeping only
the most important terms.
Taking
only terms that are leading order in $\tau$ into account, we find
\begin{equation}
F_0(0) = -i\frac{\tau^2}{3(2\pi)^3}\left(\frac{k_F}{v_F}\right)^2
\int_0^{2\pi}\frac{d\theta}{2\pi}
[1+2\cos(\theta)]\left|U_{12}(k_F\sqrt{2+2\cos(\theta)})\right|^2
\end{equation}

\section{Discussion and Conclusions}

In this paper, we have presented a fully microscopic theory of the
Coulomb drag, based on the Kubo formalism.  We have used the
finite-temperature formalism to obtain expressions for the dc drag,
and the zero-temperature formalism to obtain finite frequency results.
We have chosen to present only formal results here, deferring
the presentation of experimental consequences of these results
to another publication.\cite{KarstenBenlong}

We calculate the transconductivity $\sigma_{12}$ using an order by order
expansion in the interlayer interaction $U(\mbox{\boldmath $q$})$.
Assuming no correlations between the impurities between the
two layers, we find an exact relation between the first order
result $\sigma_{12}^{(1)}(\mbox{\boldmath $q$},\omega)$, and the subsystem
conductivities.  The result also indicates that in a uniform system,
the dc transconductivity vanishes to first order.

To second order, we write a formal result for the transconductivity
$\sigma^{(2)}_{12}$ in terms of the
$\Delta(\mbox{\boldmath $q$},\mbox{\boldmath $q$}';\omega,\omega')$-functions,
which are the
thermal-averaged $\langle j\rho\rho\rangle$ correlation functions of
the individual subsystems.
In evaluating $\Delta$ under various circumstances, we find
(i) for {\sl constant} intralayer elastic scattering rates,
we duplicate in the limit $1/\tau\rightarrow 0$
results obtained earlier using the Boltzmann equation,
and the memory functional method in the
diffusive limit; (ii) for {\sl energy-dependent} intralayer elastic
scattering rates, however, the $q^\alpha \mbox{Im}[\chi]$ must be replaced
by another quantity $F^\alpha(\mbox{\boldmath $q$},\omega)$,
which we call the transport polarizability.
The energy-dependent result is due (from the Boltzmann equation point
of view) to the the fact that the perturbed distribution function
on application of the electric field for energy-dependent elastic
scattering rates is not a drifted Fermi-Dirac.\cite{KarstenBenlong}
However, intralayer electron--electron interactions
tend to relax the distribution function back to a drifted Fermi-Dirac,
and hence the larger the intralayer e--e interactions are, the closer
the $F^\alpha(\mbox{\boldmath $q$},\omega)$ will be to $q^\alpha
\mbox{Im}[\chi_0]$ in
Eq.\ (\ref{rho21}).\cite{KarstenBenlong}

We have calculated the weak-localization correction to the second
order transconductivity, and find that
$\delta \sigma_{21}^{WL}/\sigma_{21}^0  =
\delta\sigma_{11}^{WL}/\sigma^0_{11} +
\delta\sigma_{22}^{WL}/\sigma^0_{22}$, which implies that, to lowest
order in $(k_F\ell)^{-1}$, the {\sl transresistivity} $\rho_{21}$
is {\sl unaffected} by the weak-localization corrections.

The zero-temperature formalism indicates that the dc-drag vanishes in an
open system.  This result is reproduced in the $T\rightarrow 0$
of the finite temperature formalism, since the
$\partial_{\omega}n_{\mathrm B}(\omega)$ term in
Eq.\ (\ref{sigma21}) vanishes as $T\rightarrow 0$, and one can show
that $\Delta$ is linear in $\omega$ as  $\omega \rightarrow 0$.
This statement is valid for open systems in which are connected to
dissipative leads, and not for closed systems.\cite{rojo92}

For finite frequencies we have evaluated the leading contribution
to the transconductivity at zero temperature and found that
in the clean limit $\sigma_{21} \sim \Omega^2$.
Including disorder we showed that frequency in this expression
is replaced by $1/\tau$ and $\sigma_{21} \sim \Omega/\tau$.

The formalism that we have developed in this paper can be applied to
many different physical realizations of coupled electron systems.
It is thus straightforward to extend the calculation to include
magnetic field and work is in progress in this direction.\cite{bons}
The present formalism also forms a useful starting point for
the study of e.g. higher order intralayer interactions,
phonon mediated intralayer interaction, and
correlations caused by strong interlayer electron-electron
interactions.

\acknowledgments

We thank Martin B\o nsager,
D. E. Khmelnitskii, P. A. Lee, and Lu Sham for useful
comments and discussions.  KF was supported by the Carlsberg Foundation.

{\it Note} -- Just prior to submission of this paper,
we became aware of a preprint by Alex Kamenev
and Yuval Oreg on the same subject.

\appendix
\section{Branch cuts for the three body correlation function}
\label{app:branchcuts}

Consider the Fourier transform in time according to
Eq. (\ref{Deltafourierdef}) (we set  $\tau''=0$ since
$\Delta$ is only a function of $\tau-\tau''$ and $\tau'-\tau''$)
\begin{eqnarray}
&&\Delta(i\Omega_n+i\omega_n,i\omega_n)=
\int_0^\beta d\tau\int_0^\beta d\tau'\,
e^{i\Omega_n\tau+i\omega_n\tau'}
\Delta(xx'x'';\tau\tau')\nonumber\\
&&=\int_0^\beta d\tau\, e^{i\Omega_n\tau}\left(\int_0^\tau d\tau'\,
e^{i\omega_n\tau'}
\langle J(x\tau) \rho(x'\tau') \rho(x'')\rangle +
\int_\tau^\beta d\tau' \,e^{i\omega_n\tau'}
 \langle \rho(x'\tau')J(x\tau)\rho(x'')\rangle
\right).\label{DA}
\end{eqnarray}
We now  insert the identity $1=\sum_n |n\rangle\langle n|$,
where $\{|n\rangle\}$ is a set of eigenstates for the Hamiltonian,
between the operators in Eq.\ (\ref{DA}). After
performing the two imaginary time integrals and some algebra we obtain
\begin{eqnarray}
&&\Delta(i\Omega_n+i\omega_n,i\omega_n)=
e^{\beta\Omega}
\sum_{kml}\left(\langle k|J(x)| m \rangle\langle m | \rho(x')|l\rangle
\langle l|\rho(x'')|k\rangle \frac{1}{i\omega_n + E_m -E_l}
\right.\nonumber\\
&&\times\left\{ \frac{1}{i\Omega_n+i\omega_n+E_k-E_l}
\left(e^{-\beta E_l}-e^{-\beta E_k}\right)+
\frac{1}{i\Omega_n+E_k-E_m}\left(e^{-\beta E_k}-e^{-\beta E_m}\right)
\right\}\nonumber\\
&&+\langle m|J(x)| k \rangle\langle l | \rho(x')|m\rangle
\langle k|\rho(x'')|l\rangle \frac{1}{i\omega_n + E_l -E_m}\nonumber\\
&&\times\left.\left\{ \frac{1}{i\Omega_n+i\omega_n+E_l-E_k}
\left(e^{-\beta E_l}-e^{-\beta E_k}\right)+
\frac{1}{i\Omega_n+E_m-E_k}\left(e^{-\beta E_k}-e^{-\beta E_m}\right)
\right\}\right)\label{Deltabranch}
\end{eqnarray}
{}From this expression we can read of the branch cuts of
$\Delta(i\Omega_n+i\omega_n,i\omega_n)$, which is used
when doing the Matsubara sum over $i\omega_n$ in Eq.\ (\ref{S}).

\section{Proof of $\Delta(++)=0$}
\label{app:zero}

We consider the quantity
\begin{equation}
\bar{\Delta}({\mbox{\boldmath $x$}'},i\Omega_n+i\omega_n,i\omega_n)
=\int_0^\beta d\tau'\int_0^\beta d\tau\,e^{i\Omega_n\tau}
e^{i\omega_n\tau'}
\langle T(J(\tau)\rho({\mbox{\boldmath $x$}'}\tau')\rho(0))\rangle,
\end{equation}
where $J = \int d{\mbox{\boldmath $x$}} j_\mu({\mbox{\boldmath $x$}})$. The
Fourier of
$\bar{\Delta}({\mbox{\boldmath $x$}'},i\Omega_n+i\omega_n,i\omega_n)$ with
respect to ${\mbox{\boldmath $x$}'}$ is
$\Delta(\mbox{\boldmath $q$},\mbox{\boldmath
$q$};i\Omega_n+i\omega_n,i\omega_n)$.

We need the two combinations $\Delta(+-)$ and $\Delta(++)$.
It is clear that if we set $i\Omega_n=0$ in expression above
we lose the information necessary to evaluate $\Delta(+-)$
and we can only get $\Delta(++)$ by the substitution
$i\omega_n\rightarrow \omega+i\delta$.  The dc-limit of
$\Delta(++)$ can be safely evaluated by setting
$i\Omega_n=0$ before the analytic continuation.
Doing that we obtain
\begin{equation}
\bar{\Delta}({\mbox{\boldmath $x$}'},i\omega_n,i\omega_n)
=\beta\int_0^\beta d\tau'\, e^{i\omega_n\tau'}
\langle T(J(\tau)\rho({\mbox{\boldmath $x$}'}\tau')\rho(0))\rangle.
\end{equation}
Now consider the density-density correlation function
for the case when the Hamiltonian has been
enlarged by a vector potential term
\begin{equation}
H \rightarrow H + A_\mu J_\mu.
\end{equation}
(Omitting a diamagnetic term which is not important for present argument.)
If we view  $A_\mu$ as a perturbation
we can write the charge-charge correlation function as
\begin{equation}
\chi_{A}({\mbox{\boldmath $x$}'},\tau')
= \langle T(\exp\left(-\int_0^\beta d\tau A_\mu J_\mu(\tau)\right)
\rho({\mbox{\boldmath $x$}'}\tau')\rho(0))\rangle.
\end{equation}
{}From this expression it is seen that
the function $\bar{\Delta}$ can be obtained as
\begin{equation}
\bar{\Delta}_\mu({\mbox{\boldmath $x$}'},\tau') =
-\left.\frac{d\chi_{A}({\mbox{\boldmath $x$}'},\tau')}{d A_\mu}
\right|_{A_\mu=0}.
\end{equation}
Since a constant vector potential
can always be removed by a gauge transformation
(this is however only strictly true for a system with open boundary
conditions),
$\chi_{A}$ can not depend on $A$, and hence we arrive
at the conclusion that $\bar{\Delta}=0$.

As a specific example we now take the impurity averaged
$\Delta(++)$-function in the simplifying case where we can
neglect the charge vertex corrections (i.e.,
not in the diffusive limit). Setting $i\Omega_n=0$ in Eq.\ (\ref{calH}),
we then have
\begin{eqnarray}
\frac{k^\alpha}{m}{\mbox{$\cal K$}}({\mbox{\boldmath $k$}},{\mbox{\boldmath
$q$}},ik_m,i\omega_n)&=&
[{\mbox{$\cal G$}}({\mbox{\boldmath $k$}},ik_m)]^2
\frac{k^\alpha}{m}\gamma({\mbox{\boldmath $k$}};ik_m)
{\mbox{$\cal G$}}({\mbox{\boldmath $k$}}+{\mbox{\boldmath $q$}},ik_m+i\omega_n)
\nonumber\\
&=&
-\frac{\partial {\mbox{$\cal G$}}({\mbox{\boldmath $k$}},ik_m)}{\partial
k^\alpha}
{\mbox{$\cal G$}}({\mbox{\boldmath $k$}}+{\mbox{\boldmath
$q$}},ik_m+i\omega_n),
\nonumber
\end{eqnarray}
where we have used the Ward identity: $\partial_{\mbox{\boldmath $k$}}
G^{-1}(\mbox{\boldmath $k$},ik_m) = \mbox{\boldmath $k$}\gamma(\mbox{\boldmath
$k$},ik_m)/m$.
Integrating by parts, we obtain
\begin{eqnarray}
\Delta^{\alpha}(\mbox{\boldmath $q$},\mbox{\boldmath $q$},;i\omega_n,i\omega_n)
&=&\frac{\partial}{\partial q^\alpha}\frac{e}{m} \frac{1}{\beta}
\sum_{ik_m}\int \frac{d\mbox{\boldmath $k$}}{(2\pi)^2}
\bigl({\mbox{$\cal G$}}(\mbox{\boldmath $k$},ik_m)
{\mbox{$\cal G$}}(\mbox{\boldmath $k$}+\mbox{\boldmath $q$},ik_m+i\omega_n)
\nonumber\\
&&-{\mbox{$\cal G$}}(\mbox{\boldmath $k$},ik_m)
{\mbox{$\cal G$}}(\mbox{\boldmath $k$}
-\mbox{\boldmath $q$},ik_m-i\omega_n)\Bigr)=0,
\end{eqnarray}
which can be seen by shifting $ik_n$ and $\mbox{\boldmath $k$}$ in the term.

\section{Imaginary part of the polarizability in the
diffusive limit}
\label{app:impartpol}

In this appendix, we derive expressions for Im$[\chi]$ when
$q v_F \tau < 1$ and $\omega\tau <1$; i.e, in the diffusive limit.
The polarizability including the vertex correction $\Gamma$ is given by
\begin{equation}
\chi(q,i\omega_n) = \frac{2}{\nu\beta} \sum_{\mbox{\boldmath $k$}}
\sum_{i\varepsilon_n}
G(\mbox{\boldmath $k$}+{\mbox{\boldmath $q$}},i\varepsilon_n + i\omega_n)
G(\mbox{\boldmath $k$},i\varepsilon_n)
\Gamma(q;i\varepsilon_n+i\omega_n,i\varepsilon_n).
\label{app3:1}
\end{equation}
For $qv_F\tau<1$ and $\omega\tau<1$, $\Gamma$
is given by Eq. (\ref{Gammaladder}).
Inserting this form of the vertex correction in Eq.\ (\ref{app3:1}),
writing the sum as a contour integral and deforming the contour
in the standard manner\cite{mahan} yields,
\begin{eqnarray}
\chi(q,\omega) &=& -\frac{2}{\nu}\sum_k
\int \frac{d\varepsilon}{2\pi i}\ n_F(\varepsilon) \Bigl\{
[\Gamma^-(q,\omega) G^r(\mbox{\boldmath $k$}+{\mbox{\boldmath
$q$}},\varepsilon)
-G^a(\mbox{\boldmath $k$}+{\mbox{\boldmath $q$}},\varepsilon)]
G^a(\mbox{\boldmath $k$},\varepsilon - \omega)\nonumber\\
&&+
[G^r(\mbox{\boldmath $k$},\varepsilon) - \Gamma^-(q,\omega)
G^a(\mbox{\boldmath $k$},\varepsilon)]
G^r(\mbox{\boldmath $k$}+{\mbox{\boldmath $q$}},\varepsilon + \omega)
\Bigr\}.
\label{app3:2}
\end{eqnarray}
In the above equation, the analytic continuation $i\omega_n\rightarrow
\omega + i0^+$ has been taken, and
$\Gamma^\pm(q,\omega)$ is defined in Eq. (\ref{Gammapm}).

We write $\chi = \chi_a + \chi_b$, where
$\chi_a$ ($\chi_b$) excludes (includes) the vertex corrections.
The term $\chi_a$ to lowest order in $q$ and $\omega$ is
\begin{eqnarray}
\chi_a(q,\omega) &\equiv& -\frac{2}{\nu}\sum_k \int \frac{d\varepsilon}
{2\pi i} n_F(\varepsilon) [G^r(\mbox{\boldmath $k$},\varepsilon)
G^r(\mbox{\boldmath $k$}+{\mbox{\boldmath $q$}},\varepsilon + \omega) -
G^a(\mbox{\boldmath $k$}+{\mbox{\boldmath $q$}},\varepsilon)
G^a(\mbox{\boldmath $k$},\varepsilon - \omega)]
\nonumber\\
&\approx&\frac{2}{\nu}\sum_k\int \frac{d\varepsilon}{2\pi i}
n_F(\varepsilon) \frac{\partial [G^r(\mbox{\boldmath $k$},\varepsilon) -
G^a(\mbox{\boldmath $k$},
\varepsilon)]}{\partial\varepsilon}\nonumber\\
&=&-\int \frac{d\varepsilon}{2\pi i}
n_F'(\varepsilon)\frac{2}{\nu}\sum_k
[G^r(\mbox{\boldmath $k$},\varepsilon) - G^a(\mbox{\boldmath $k$},\varepsilon)]
= - \frac{\partial n}{\partial \mu}
\end{eqnarray}
Note that $\chi_a$ to lowest order in $q$ and $\omega$ is purely real.

The second part is
\begin{equation}
\chi_b(q,\omega) = -\Gamma^-(q,\omega) \frac{2}{\nu} \sum_k
\int \frac{d\varepsilon}{2\pi i} n_F(\varepsilon)
[G^r(\mbox{\boldmath $k$}+{\mbox{\boldmath
$q$}},\varepsilon)G^a(\mbox{\boldmath $k$},\varepsilon - \omega)
- G^a(\mbox{\boldmath $k$},\varepsilon) G^r(\mbox{\boldmath
$k$}+{\mbox{\boldmath $q$}},\varepsilon + \omega)].
\label{impol}
\end{equation}
To lowest order in $q$ and $\omega$,
this is\cite{alts80}
\begin{equation}
\chi_b(q,\omega) = \frac{\partial  n}{\partial \mu}
\frac{-i\omega}{Dq^2 -i\omega}.
\end{equation}
Thus $\chi_b$ has an imaginary component which, as shown in
section \ref{sec:imp},
is related to the $\Delta$ when $qv_F\tau <1$ and $\omega\tau <1$.

\section{Justification for neglecting $G^rG^r$- and $G^aG^a$-terms}
\label{app:proof}

In this appendix, we show that the terms in $\Delta_L$
involving products $G^r G^r$ and $G^a G^a$ (denoted below
as $\Delta_s$) can be neglected
when compared to $G^a G^r$ when $E_F\tau\gg 1$, in the
diffusive limit.

Written in full, $\Delta_s$ is
\begin{eqnarray}
\Delta_s^\alpha(q,q;\omega'+i\delta,\omega'-i\delta) &=&
-\frac{2\tau}{m} \frac{1}{\nu} \sum_k
\int \frac{d\varepsilon}{2\pi}
n_F(\varepsilon) k^\alpha\nonumber\\\
&&\Biggl\{
\Gamma^-(q,\omega)\Bigl[
-G^r({\mbox{\boldmath $k$}},\varepsilon)
G^r({\mbox{\boldmath $k$}}+{\mbox{\boldmath $q$}},\varepsilon+\omega)
-G^a({\mbox{\boldmath $k$}},\varepsilon)
G^a({\mbox{\boldmath $k$}}-{\mbox{\boldmath
$q$}},\varepsilon-\omega)\nonumber\\
&& +G^r({\mbox{\boldmath $k$}},\varepsilon-\omega)
G^r({\mbox{\boldmath $k$}}+{\mbox{\boldmath $q$}},\varepsilon)
+G^a({\mbox{\boldmath $k$}},\varepsilon+\omega)
G^a({\mbox{\boldmath $k$}}-{\mbox{\boldmath $q$}},\varepsilon)
\Bigr] + c.c \Biggr\}
\end{eqnarray}
Note that at $\omega=0$, this term is identically zero.
Expanding in powers of $\omega$ gives
\begin{eqnarray}
\Delta_s^\alpha(q,q;\omega'+i\delta,\omega'-i\delta) &=&
\mbox{Re}\Biggl[
-\omega \Gamma^-(q,\omega)\frac{\tau}{m} \frac{4}{\nu} \sum_k
\int \frac{d\varepsilon}{2\pi} n_F(\varepsilon)\, k^\alpha \nonumber\\
&& \frac{ \partial
[G^a({\mbox{\boldmath $k$}},\varepsilon) G^a({\mbox{\boldmath
$k$}}-{\mbox{\boldmath $q$}},\varepsilon)
-G^r({\mbox{\boldmath $k$}},\varepsilon) G^r({\mbox{\boldmath
$k$}}+{\mbox{\boldmath $q$}},\varepsilon)]}
{\partial \varepsilon} + O(\omega^2)
\Biggr] \nonumber\\
&=&
\mbox{Re}\Biggl[-
\omega \Gamma^-\frac{\tau}{m} \frac{4}{\nu} \sum_k
\int \frac{d\varepsilon}{2\pi} n_F'(\varepsilon) k^\alpha\nonumber\\
&&
[G^a({\mbox{\boldmath $k$}},\varepsilon) G^a({\mbox{\boldmath
$k$}}-{\mbox{\boldmath $q$}},\varepsilon)
-G^r({\mbox{\boldmath $k$}},\varepsilon) G^r({\mbox{\boldmath
$k$}}+{\mbox{\boldmath $q$}},\varepsilon)]
\Biggr] + O(\omega^2).
\label{app4:2}
\end{eqnarray}
Expanding Eq.\ (\ref{app4:2}) in powers of $q$, and assuming that
the self-energy is small (since $\tau$ is large)
so that $\partial G/\partial \varepsilon_{\mbox{\boldmath $k$}} \approx G^2$,
yields
\begin{eqnarray}
\Delta_s^\alpha(q,q;\omega'+i\delta,\omega'-i\delta) &=&
\mbox{Re}\Biggl[ \omega q^\alpha \Gamma^-(q,\omega)
\frac{\tau}{m} \frac{4}{\nu} \sum_k
\int \frac{d\varepsilon}{2\pi}
n_F'(\varepsilon)\frac{(k^\alpha)^2}{m}\nonumber\\
&&[G^a({\mbox{\boldmath $k$}},\varepsilon)^3 + G^r({\mbox{\boldmath
$k$}},\varepsilon)^3]
\Biggr] + O(q^2,\omega^2).
\end{eqnarray}
Integrals of $G^{a,r}({\mbox{\boldmath $k$}},\varepsilon)^3$ over
${\mbox{\boldmath $k$}}$
do {\sl not} diverge when $\tau\rightarrow\infty$ because the poles
of the function are on the same half-plane (unlike integrals over
$G^a({\mbox{\boldmath $k$}},\varepsilon) G^r({\mbox{\boldmath
$k$}},\varepsilon)$, which
go as $\tau$).   Since $n_F'(\varepsilon)$ is peaked around $\mu$,
one can estimate the magnitude of $\Delta_s^\alpha$ by replacing
$-n_F'(\varepsilon) \approx \delta(\varepsilon-\mu)$, which gives
\begin{equation}
\Delta_s^\alpha(q,q';\omega'+i\delta,\omega'-i\delta) \approx \mbox{Re}
\Biggl[\Gamma^-(q,\omega)\frac{e\tau q^\alpha \omega}{
2\pi^3 E_F}\Biggr].
\end{equation}
A comparison of this with $\Delta_L^\alpha$ given in Eq. (\ref{deleqchi})
shows that $\Delta_s^\alpha$ is smaller by a factor of $1/(E_F\tau)$.

%\bibliographystyle{prsty}
%\bibliography{/users/karsten/tex/karsten}

\begin{thebibliography}{10}
\bibitem{non-trans}
For a system where the two subsystems scatter from the same impurities,
  diagrams that connect the two conductivity bubbles must be included and the
  transconductivity cannot be written as a product of the impurity averaged
  conductivities of the two subsystems. See also footnote
  \protect\onlinecite{imp-imp}.
\bibitem{6-lines}
It is easy to verify (\ref{1storder}) independently by the following simple
  argument, which takes into account the external field experienced by layer 2
  due to charge perturbations in layer 1, \begin{eqnarray}
  (-e)j_2^\alpha({\mbox{\boldmath $q$}},\omega) &=& \sum_{\delta}
  \sigma_{22}^{\alpha\delta}({\mbox{\boldmath $q$}},\omega)
  E_2^{\delta}({\mbox{\boldmath $q$}},\omega) \nonumber\\ &=&
  \sum_{\delta}{\mbox{\boldmath $q$}^{\delta}
  \over{i}}\sigma_{22}^{\alpha\delta} ({\mbox{\boldmath $q$}},\omega) {1\over
  {(-e)}} U_{12}({\mbox{\boldmath $q$}},\omega)\rho_1({\mbox{\boldmath
  $q$}},\omega)\nonumber\\ &=& \sum_{\gamma,\delta}{{\mbox{\boldmath
  $q$}}^{\delta}\over{-ie}} \sigma_{22}^{\alpha\delta}({\mbox{\boldmath
  $q$}},\omega) U_{12}({\mbox{\boldmath $q$}},\omega){{\mbox{\boldmath
  $q$}}^{\gamma}\over{\omega}} j_1^{\gamma}({\mbox{\boldmath
  $q$}},\omega)\nonumber\\ &=& \sum_{\beta,\gamma,\delta}\left[
  {{q^{\delta}q^{\gamma}}\over{ie^2\omega}}
  \sigma_{22}^{\alpha\delta}({\mbox{\boldmath $q$}},\omega)
  U_{12}({\mbox{\boldmath $q$}},\omega)
  \sigma_{11}^{\gamma\beta}({\mbox{\boldmath $q$}},\omega)
  \right]E_{1\text{ext}}^{\beta}({\mbox{\boldmath $q$}},\omega), \nonumber
  \end{eqnarray} where we used continuity equation to relate $\rho_1$ and
  $\mbox{\boldmath $j$}_1$. The charge of the carriers was taken to be $-e$,
  although a generalization to mixed carrier systems is straightforward. The
  term in the square brackets is the transconductivity tensor, which agrees
  with Eq.\ (\ref{1storder}).
\bibitem{crossed}
One should note that Fig. \ref{fig:vertex} does not include all possible
  diagrams. Examples of such higher-order diagrams are analyzed in Section
  \ref{sec:WL}.
\end{thebibliography}

\begin{figure}
\caption{\label{fig:crossed}
The different types of crossed diagrams for the
triangle diagram relevant for the weak localization
correction to the transconductivity.
Diagrams of the type in (a) give the leading order
contribution. Dressing the charge vertexes as in (b)
gives a smaller contribution for moderately clean
samples for the same reason that allows us to
neglect the vertex corrections of the charge
vertexes, discussed in Sec.\ \protect\ref{sec:ladder}.
The diagram in (c), which cannot be included using vertex functions
alone, has an even smaller phase space
and can hence also be neglected.}
\end{figure}

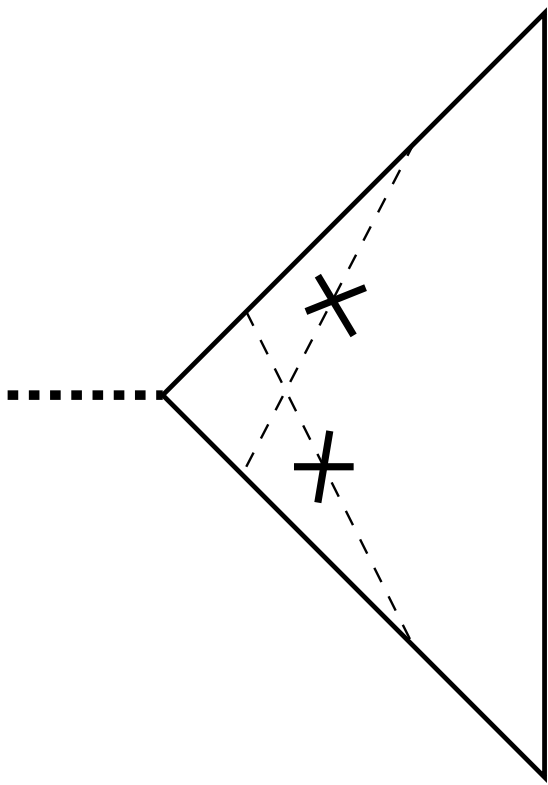
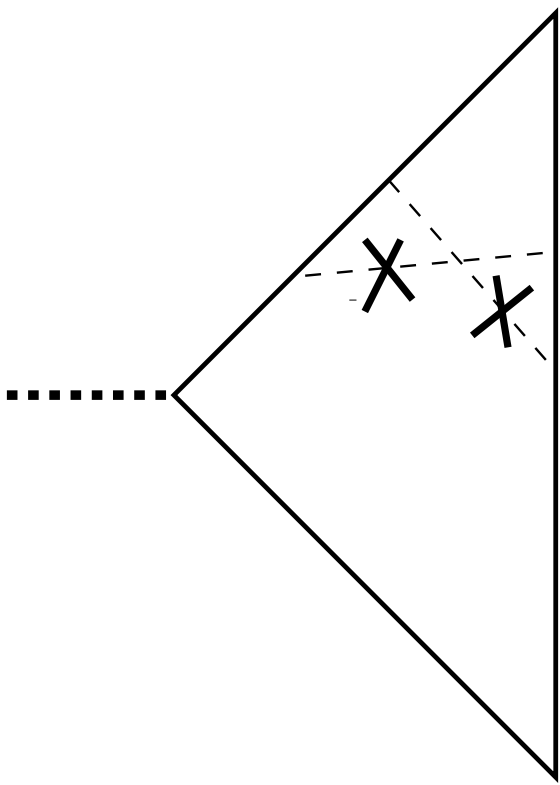
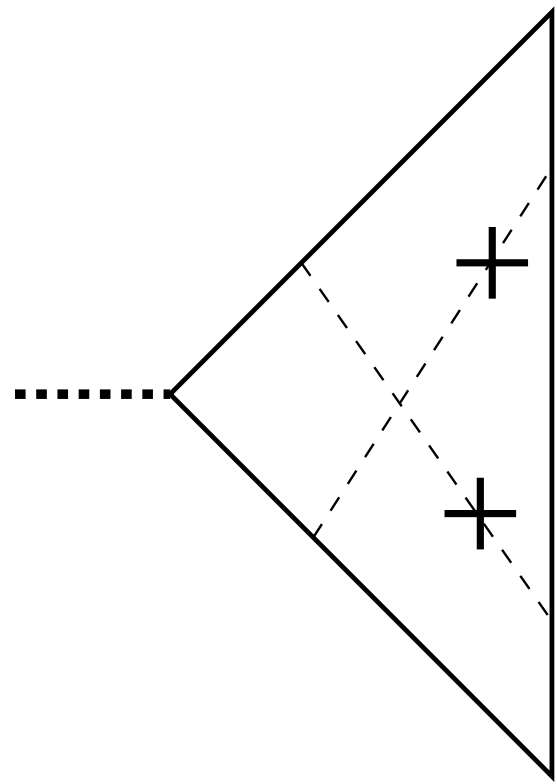
\end{document}